\documentclass[11pt]{article}
\usepackage{amssymb}

\usepackage[totalwidth=460truept,totalheight=600truept]{geometry}
\usepackage{epsfig,latexsym}

\def\theequation{\arabic{section}.\arabic{equation}}

\renewcommand{\theequation}{\thesection.\arabic{equation}}
\global\arraycolsep=1truept
\input epsf
\input{tcilatex}
\begin{document}

\font\cmss=cmss10 \font\cmsss=cmss10 at 7pt \hfill \hfill IFUP-TH 2005/08


\vskip 1.4truecm

\begin{center}
{\huge \textbf{Infinite Reduction Of Couplings In}}

\vskip .4truecm

{\huge \textbf{Non-Renormalizable Quantum Field Theory}}

\vskip 1.5truecm

\textsl{Damiano Anselmi}

\textit{Dipartimento di Fisica ``Enrico Fermi'', Universit\`{a} di Pisa, }

\textit{Largo Bruno Pontecorvo 3, I-56127 Pisa, Italy, }

\textit{and INFN, Sezione di Pisa, Italy}

e-mail: anselmi@df.unipi.it
\end{center}

\vskip 2truecm

\begin{center}
\textbf{Abstract}
\end{center}

{\small I study the problem of renormalizing a non-renormalizable theory
with a reduced, eventually finite, set of independent couplings. The idea is
to look for special relations that express the coefficients of the
non-renormalizable terms as unique functions of a reduced set of independent
couplings }$\lambda ${\small , such that the divergences are removed by
means of field redefinitions plus renormalization constants for the }$%
\lambda ${\small s. I consider non-renormalizable theories whose
renormalizable subsector }$\mathcal{R}${\small \ is interacting. The
``infinite'' reduction is determined by i) perturbative meromorphy around
the free-field limit of }$\mathcal{R}${\small , or ii) analyticity around
the interacting fixed point of }$\mathcal{R}${\small . In general,
prescriptions i) and ii) mutually exclude each other. When the reduction is
formulated using i), the number of independent couplings remains finite or
slowly grows together with the order of the expansion. The growth is slow in
the sense that a reasonably small set of parameters is sufficient to make
predictions up to very high orders. Instead, in case ii) the number of
couplings generically remains finite. The infinite reduction is a tool to
classify the non-renormalizable interactions and address the problem of
their physical selection.}

\vskip1truecm

\vfill\eject

\section{Introduction}

\setcounter{equation}{0}

Fundamental theories should be able, at least in principle, to describe
arbitrarily high energies with a finite number of independent couplings. The
usual formulation of non-renormalizable theories makes use of infinitely
many independent couplings to subtract the divergences. So formulated,
non-renormalizable theories are good only as effective field theories,
finitely many parameters being sufficient to make predictions about
low-energy phenomena.

These facts, however, do not imply that non-renormalizable theories are
useless or inadequate as fundamental theories, but only that their naive
formulation is. An improved formulation should be uncovered. Some steps in
this direction have been made in ref.s \cite{pap2,pap3,pap6}. In \cite
{pap2,pap3} finite and quasi-finite non-renormalizable theories have been
constructed as irrelevant deformations of interacting conformal field
theories. In \cite{pap6} a certain class of non-renormalizable theories with
a running renormalizable subsector $\mathcal{R}$ have been studied, in a
perturbative framework of new type, which allows to treat unexpanded
functions of the fields, although not of their derivatives. In some of the
models of \cite{pap6}, with non-analytic potentials, the divergences are
reabsorbed with a finite number of independent couplings. In more standard
models the number of independent couplings grows together with the order of
the expansion, but a certain form of predictivity is retained.

The purpose of this paper is to generalize the results of \cite{pap6},
develop the systematics of the ``reduction of couplings'' for
non-renormalizable theories and study the predictive power of the reduced
theories. The reduction of couplings is the search for special, unambiguous
and self-consistent relations among the couplings, such that the lagrangian
depends on a reduced, eventually finite, set of couplings $\lambda $ and all
divergences are removed by means of field redefinitions plus renormalization
constants for the $\lambda $s. The reduction is a tool to ``diagonalize'',
and therefore classify, the non-renormalizable interactions. The potential
applications of this investigation are to physics beyond the Standard Model
and quantum gravity.

\bigskip

Unless otherwise specified, the words ``relevant'', ``marginal'' and
``irrelevant'' refer to the Gaussian fixed point, so they are equivalent to
``super-renormalizable'', ``strictly renormalizable'' and
``non-renormalizable'', respectively. In the study of deformations of
interacting conformal field theories, the construction of this paper allows
also to characterize the deformation as marginal, relevant or irrelevant at
the interacting fixed point.

The power-counting renormalizable sector $\mathcal{R}$ needs to be fully
interacting, by which I mean that all marginal interactions are turned on.
Indeed, the infinite reduction does not work when the marginal sector is
free or only partially interacting. Without loss of generality, I assume
also that $\mathcal{R}$ does not contain relevant couplings. This assumpion
ensures that the beta functions depend polynomially on the irrelevant
couplings. When $\mathcal{R}$ is fully interacting, relevant parameters can
be added perturbatively \textit{after} the construction of the irrelevant
deformation.

The inclusion of relevant parameters with a free or only partially
interacting renormalizable sector $\mathcal{R}$ is important for
applications to quantum gravity in four dimensions, which has no marginal
coupling, its relevant parameter being by the cosmological constant.
However, further insight is needed to deal with the technical complicacies
of this problem, so its investigation is postponed. Basically, in the
constructions of this paper the interactions have to be turned on in the
following order: first the marginal interactions, then the irrelevant
interactions, finally the relevant interactions.

Denote the marginal couplings of $\mathcal{R}$ with $\alpha $ and the
irrelevant couplings of the complete theory with $\lambda _{n}$. The
subscript $n$ denotes the ``level'' of $\lambda _{n}$, $-n$ being the
dimensionality of $\lambda _{n}$ in units of mass. The beta functions of the
irrelevant couplings have the structure 
\begin{equation}
\beta _{\lambda _{n}}(\alpha ,\lambda )=\gamma _{n}\left( \alpha \right)
\lambda _{n}+\delta _{n}(\lambda _{m<n},\alpha ),  \label{e1}
\end{equation}
where $\delta _{n}(\lambda _{m<n},\alpha )$ is polynomial, at least
quadratic, in the irrelevant couplings $\lambda _{m}$ with $m<n$. The
structure (\ref{e1}) is obtained matching the dimensionalities of the left-
and right-hand sides. Indeed, in perturbation theory only integer powers of
the couplings can appear and by assumption there are no couplings with
positive dimensionalities in units of mass. Therefore $\beta _{\lambda _{n}}$
is at most linear in $\lambda _{n}$, polynomial in the irrelevant couplings $%
\lambda _{k}$ with $k<n$ and does not depend on the irrelevant couplings $%
\lambda _{k}$ with $k>n$. It is convenient to separate the $\lambda _{n}$%
-independent contributions, collected in $\delta _{n}$, from those that are
proportional to $\lambda _{n}$. All monomials $\prod_{k<n}\lambda
_{k}^{n_{k}}$ contained in $\delta _{n}$ satisfy $\sum_{k<n}kn_{k}=n$, so
they are at least quadratic in $\lambda _{k}$ with $k<n$. Of course, $\beta
_{\lambda _{n}}$ can depend non-polynomially on the marginal couplings $%
\alpha $.

An irrelevant deformation is made of a head and a queue. The head is the
lowest-level irrelevant term. Denote its coupling with $\overline{\lambda }$%
. The queue is made of the other irrelevant terms, whose couplings $\lambda
_{n}$ are not independent, but unique functions of $\overline{\lambda }$ and 
$\alpha $, given by certain ``reduction relations'' $\lambda _{n}=\lambda
_{n}(\alpha ,\overline{\lambda })$, to be determined. Differentiating the
reduction relations with respect to the dynamical scale $\mu $, the RG
consistency conditions 
\begin{equation}
\beta _{\lambda _{n}}(\alpha ,\lambda )-\frac{\partial \lambda _{n}}{%
\partial \overline{\lambda }}\beta _{\overline{\lambda }}=\frac{\partial
\lambda _{n}}{\partial \alpha }\beta _{\alpha }  \label{e2}
\end{equation}
are obtained. The consistency conditions (\ref{e2}) ensure that the
divergences of the theory are renormalized just by the renormalization
constants of $\alpha $ and $\overline{\lambda }$, plus field redefinitions.
Nevertheless, (\ref{e2}) are not sufficient to determine the reduction
relations uniquely, because their solutions contain arbitrary finite
parameters $\xi _{n}$. Extra assumptions have to be introduced to have a
true reduction.

In the realm of power-counting renormalizable theories similar problems were
first considered by Zimmermann \cite{zimme,oheme}, who suggested to
eliminate the $\xi $-ar\-bi\-tra\-ri\-ness requiring that the reduction
relations be analytic, for consistency with perturbation theory. The
analytic reduction works in a set of models, when the reduced theory
contains a single independent coupling, but is problematic when the reduced
theory contains more than one independent coupling \cite{wilson}.

Zimmermann's approach can be understood as an alternative to unification.
Its phenomenological implications have been investigated for example in \cite
{pheno}. For a technical review, see \cite{review}. It is also possible to
use Zimmermann's method to construct finite N=1 supersymmetric theories \cite
{lucchesi}. Beyond power-counting, Zimmermann's approach has been studied by
Atance and Cortes in effective scalar theories and quantum gravity \cite
{cortes1,cortes2} and by Kubo and M. Nunami \cite{giap} using the Wilsonian
approach, but the systematics of the reduction of couplings in
non-renormalizable theories (which I\ call \textit{infinite reduction}) has
not been developed, so far.

\bigskip

In the infinite reduction, some issues are different than in Zimmermann's
reduction. First, note that no $\xi $-ambiguity affects the finite and
quasi-finite non-renormalizable theories of ref.s \cite{pap2,pap3}. Indeed,
the $\overline{\lambda }$-dependence of $\lambda _{n}$ is unambiguously
fixed on dimensional grounds: $\lambda _{n}=$ $\overline{\lambda }^{n/\ell
}f_{n}(\alpha )$, where $\ell $ is the level of $\overline{\lambda }$. So,
when the renormalizable sector $\mathcal{R}$ is a conformal field theory $%
\mathcal{C}$ ($\beta _{\alpha }=0$) the differential equations (\ref{e2})
collapse into algebraic equations. Because the equations are algebraic, the
solutions do not contain new independent parameters. Because the equations
are linear in their own unknowns $\lambda _{n}$, the solution exists and is
unique, under certain conditions that are reviewed in sections 2 and 3.
Finally, because $\delta _{n}$ depends only on the irrelevant couplings $%
\lambda _{m}$ with $m<n$, the construction is algorithmic in the level $n$.

It was shown in \cite{pap2,pap3} that the free-field limit ($\alpha
\rightarrow 0$) of the deformed theory is singular in $\alpha $, and that
the maximal singularity multiplying an irrelevant operator is bounded by the
dimensionality of the operator itself, or, equivalently, by the power of $%
\overline{\lambda }$. These facts mean that: $i$) the reduction is not
analytic, but \textit{meromorphic}; $ii$) the singularity can be reabsorbed
into $\overline{\lambda }$, defining a suitable ``effective Planck mass''
for the irrelevant interaction. A meromorphy of this type, where the
negative powers can be arbitrarily high, but the maximal negative power
grows linearly with the order of some expansion is called \textit{%
perturbative meromorphy}.

Equipped with the knowledge learnt from ref.s \cite{pap2,pap3}, I study
prescriptions to remove the $\xi $-ambiguity in the irrelevant deformations
of running renormalizable theories $\mathcal{R}$. I show that the reduction
relations are uniquely determined by perturbative meromorphy around the
free-field limit, if some existence conditions are fulfilled, e.g. certain
linear combinations of one-loop anomalous dimensions, normalized with the
one-loop coefficient of the $\mathcal{R}$ beta function, do not coincide
with natural numbers. A non-trivial renormalization mixing makes the
existence conditions less restrictive. Most of the $\xi $-arbitrariness is
removed with this prescription, but sometimes the existence conditions are
violated. Then, new independent couplings have to be introduced at high
orders. In some models the number of independent couplings of the complete
theory is finite, in other models it grows together with the order of the
expansion. A form of predictivity is retained also in the latter case,
because in general the growth is slow and a reasonably small number of
parameters is sufficient to make predictions up to very high orders. Models
of this type have been studied in \cite{pap6}.

The infinite reduction is scheme-independent, because the existence
conditions involve only one-loop coefficients.

An alternative scheme-independent prescription for the infinite reduction is
analyticity around an interacting fixed point of $\mathcal{R}$. In this
case, the number of independent couplings generically remains finite in the
complete reduced theory. Nevertheless, perturbative meromorphy around the
free fixed point and analyticity around the interacting fixed point mutually
exclude each other. Similarly, when $\mathcal{R}$ interpolates between two
interacting fixed points, the reduction relations can be analytic only
around one of them at a time. These features of the infinite reduction are a
bit disappointing. However, it should be kept in mind that in the realm of
non-renormalizable theories it is meaningful to impose conditions only
around the IR fixed point, free or interacting, because the ultraviolet
limit is not required to exist.

\bigskip

The study of quantum field theory beyond power counting has attracted
interest for decades, motivated by low-energy QCD and quantum gravity. Some
non-renormalizable models can be constructed with \textit{ad hoc}
procedures, such as the large N expansion, used for three-dimensional
four-fermion theories \cite{parisi} and sigma models \cite{arefeva2}.
Weinberg's asymptotic safety \cite{wein} is a more general scenario. The
theory is assumed to have an interacting fixed point in the ultraviolet with
a finite-dimensional critical surface. The RG flow tends to the fixed point
only if the irrelevant couplings are appropriately fine-tuned. In general,
only a finite number of arbitrary parameters survives this fine tuning.
Asymptotic safety has been recently studied for gravity \cite{reuter} and
the Higgs sector of the Standard Model \cite{wette} using the ERG (exact
renormalization-group) approach.

The paper is organized as follows. In sections 2 and 3 I review and
elaborate on the finite and quasi-finite irrelevant deformations of
interacting conformal field theories \cite{pap2,pap3}. In section 4
formulate the general principles of the infinite reduction and work out the
conditions under which the number of independent parameters can be reduced
and eventually kept finite. In sections 5 I propose an interpretation of the
infinite reduction. In section 6 I study the infinite reduction around
interacting fixed points. Section 7 contains some illustrative applications.
In section 8 I\ discuss irrelevant deformations of theories that contain
more than one marginal coupling. Section 9 contains the conclusions.
Appendix A contains a brief review of Zimmermann's approach and a comparison
with the infinite reduction. Appendix B contains definition and properties
of perturbative meromorphy for the infinite reduction.

\section{Finiteness beyond power-counting}

\setcounter{equation}{0}

Consider a conformal field theory $\mathcal{C}$ of fields $\varphi $
interacting with the lagrangian $\mathcal{L}_{\mathcal{C}}[\varphi ,\alpha ]$%
, $\alpha $ denoting the marginal couplings. Let \textrm{O}$_{\lambda }$
denote a basis of ``essential'', local, symmetric, scalar, canonically
irrelevant operators constructed with the fields of $\mathcal{C}$ and their
derivatives. The essential operators are the equivalence classes of
operators that differ by total derivatives and terms proportional to the
field equations \cite{wein}. Total derivatives are trivial in perturbation
theory, while terms proportional to the field equations can be renormalized
away by means of field redefinitions, so they do not affect the beta
functions of the physical couplings. Finally, the operators \textrm{O}$%
_{\lambda }$ are Lorentz scalars and have to be ``symmetric'', that is to
say invariant under the non-anomalous symmetries of the theory, up to total
derivatives.

The irrelevant terms can be ordered according to their level. If \textrm{O}$%
_{\lambda }$ has canonical dimensionality $d_{\lambda }$ in units of mass,
then the level $\ell _{\lambda }$ of \textrm{O}$_{\lambda }$ is the
difference $d_{\lambda }-d$, $d$ being the spacetime dimension. If $\lambda $
denotes the coupling that multiplies the operator \textrm{O}$_{\lambda }$,
then $\ell _{\lambda }$ is minus the naive dimensionality of $\lambda $. It
is understood that in general each level contains finitely many operators,
which can mix under renormalization. For the moment I do not distinguish
operators of the same level. For concreteness, formulas are written in the
case $d=4$, because the generalization to other $d$'s is simple.

The lagrangian of the deformed theory reads 
\[
\mathcal{L}[\varphi ]=\mathcal{L}_{\mathcal{C}}[\varphi ,\alpha
]+\sum_{\lambda }\lambda ~\mathcal{O}_{\lambda }(\varphi ). 
\]
The beta function $\beta _{\lambda }$ of $\lambda $ has the schematic
structure (\ref{e1}) \cite{pap2} 
\begin{equation}
\beta _{\lambda }=\lambda \gamma _{\lambda }+\delta _{\lambda }.
\label{betagel}
\end{equation}
obtained matching the naive dimensionalities, where $\delta _{\lambda }$
does not depend on $\lambda $ and is polynomial, at least quadratic, in the
irrelevant couplings $\lambda ^{\prime }$ with levels $\ell _{\lambda
^{\prime }}<\ell _{\lambda }$. The coefficient $\gamma _{\lambda }(\alpha )$
of $\lambda $ is the anomalous dimension of the operator $\mathrm{O}%
_{\lambda }$, viewed as a composite operator of the undeformed theory $%
\mathcal{C}$. Operators with $\delta _{\lambda }$ equal to zero are called 
\textit{protected}. Examples of protected operators are the chiral operators
in four-dimensional supersymmetric theories \cite{grisaru}. Operators with $%
\gamma _{\lambda }=0$ are finite, as viewed from the undeformed theory $%
\mathcal{C}$.

An irrelevant deformation is made of a \textit{head} and a \textit{queue}.
The head is the first irrelevant term $\mathrm{O}_{\overline{\lambda }}$ of
the deformation, multiplied by the independent coupling $\overline{\lambda }$%
. Obviously, $\delta _{\overline{\lambda }}=0$, so the head is always
protected. The queue is made of infinitely many irrelevant terms with levels 
$\ell _{\lambda }>\ell _{\overline{\lambda }}$, multiplied by unique
functions of $\overline{\lambda }$ and $\alpha $, obtained solving the
finiteness equations 
\begin{equation}
\beta _{\lambda }=0.  \label{feq}
\end{equation}
Since $\delta _{\lambda }$ depends only on the couplings with levels $\ell
_{\lambda ^{\prime }}<\ell _{\lambda }$ the solution can be worked out
recursively in the levels $\ell _{\lambda }$.

Equation (\ref{feq}) has solutions when the operator $\mathrm{O}_{\lambda }$
is not finite ($\gamma _{\lambda }\neq 0$) and when it is finite and
protected ($\delta _{\lambda }=\gamma _{\lambda }=0$). It does not have
solutions when the operator $\mathrm{O}_{\lambda }$ is finite but not
protected. The solution is trivial ($\lambda =0$) when the operator is
protected, but not finite.

The irrelevant deformation is non-trivial if the head $\mathrm{O}_{\overline{%
\lambda }}$ is a finite operator. Indeed, recalling that the head is always
protected, the equation $\beta _{\overline{\lambda }}=0$ leaves $\overline{%
\lambda }$ arbitrary. The queue exists if it does not include any finite
unprotected operator, namely $\gamma _{\lambda }\neq 0$ any time $\delta
_{\lambda }\neq 0$. When these \textit{invertibility conditions} are
fulfilled, the couplings of the queue are recursively given by 
\begin{equation}
\lambda =-\frac{\delta _{\lambda }}{\gamma _{\lambda }}  \label{solgel}
\end{equation}
in terms of $\overline{\lambda }$ and $\alpha $.

The irrelevant deformation is trivial if the theory $\mathcal{C}$ has no
finite irrelevant operator. Indeed, in this case $\beta _{\overline{\lambda }%
}=0$ implies $\overline{\lambda }=0$ and the other finiteness conditions
iteratively imply that all $\lambda $s vanish, which gives back the
undeformed theory $\mathcal{C}$.

Summarizing, the theory $\mathcal{C}$ admits a non-trivial finite irrelevant
deformation is there exists a finite operator and no finite unprotected
operator.

The invertibility conditions are obviously violated if $\mathcal{C}$ is
free, but are expected to be generically fulfilled if $\mathcal{C}$ is fully
interacting. Examples of non-trivial finite unprotected irrelevant operators
in interacting conformal field theories are not known. The known finite
irrelevant operators, such as the chiral operators in four-dimensional
superconformal field theories, are also protected.

The anomalous dimensions $\gamma _{\lambda }$ depend on the marginal
couplings $\alpha $ of $\mathcal{C}$. It is reasonable to expect that the
anomalous dimension of an unprotected irrelevant operator vanishes at most
for special values of $\alpha $. In this sense, the requirement that the
theory $\mathcal{C}$ does not have finite irrelevant unprotected operators
is a restriction on the theory $\mathcal{C}$. Thus, in principle it is
possible to say which conformal field theories admit which deformations just
from the knowledge of the undeformed conformal theory, before turning the
irrelevant deformation on.

When some invertibility conditions are violated, the irrelevant deformation
cannot be finite, but can be easily promoted to a quasi-finite deformation
(see the next section). More precisely, when an operator $\mathcal{O}_{%
\overline{\lambda }}$ of the queue is finite and unprotected (so the
equation $\beta _{\overline{\lambda }}=0$ has no solution) it is sufficient
to treat the coupling $\overline{\lambda }$ as a new independent coupling,
free to run according to the equation $\beta _{\overline{\lambda }}=\delta _{%
\overline{\lambda }}$. The rest of the queue is then determined as explained
in the next section. These and other similar situations are discussed at
length in the paper.

Thus, it is sufficient to assume that the invertibility conditions are
violated in at most a finite number of cases to renormalize the irrelevant
deformation with a finite number of independent couplings, plus field
redefinitions.

\bigskip

Assuming that the invertibility conditions are fulfilled, the structure of
the deformed lagrangian is 
\begin{equation}
\mathcal{L}[\varphi ]=\mathcal{L}_{\mathcal{C}}[\varphi ,\alpha ]+\overline{%
\lambda }_{\ell }\mathcal{O}_{\ell }(\varphi )-\sum_{n=2}^{\infty }\frac{%
\delta _{n\ell }}{\gamma _{n\ell }}~\mathcal{O}_{n\ell }(\varphi ).
\label{def1}
\end{equation}

Now, assume that $\mathcal{C}$ is finite, namely it belongs to a
one-parameter family of conformal field theories that includes the
free-field limit. In this case, $\beta _{\alpha }(\alpha )\equiv 0$ for
every $\alpha $. Then, the deformed theory (\ref{def1}) is renormalized just
by field redefinitions, so it is finite. Instead, if $\mathcal{C}$ is the
fixed point of an RG flow ($\beta _{\alpha }(\alpha _{*})=0$ only for some
special value $\alpha =\alpha _{*}$), then ``fake'' renormalization
constants are necessary to define both $\mathcal{C}$ and (\ref{def1}),
including a non-trivial $Z_{\alpha }$. Such renormalization constants do not
cause the introduction of new physical couplings and do not affect the
renormalization-group flow at $\alpha =\alpha _{*}$. It is natural to
enlarge the notion of finiteness to include every theory of this type.

If the RG flow is defined varying the dynamical scale $\mu $ at fixed
external momenta and $\overline{\lambda }_{\ell }$, then (\ref{def1}) is a
fixed point of the flow. Instead, rescaling the overall energy $E$ of
correlations functions at fixed $\mu $ and $\overline{\lambda }_{\ell }$,
every insertion of $\int \mathcal{O}_{n\ell }$ rescales, by construction,
with canonical exponent $n\ell $. That means that the deformation (\ref{def1}
) is irrelevant not only with respect to the Gaussian fixed point, but also
with respect to the interacting conformal field theory $\mathcal{C}$.

\bigskip

Operators of the same level can be distinguished with extra indices in $%
\lambda $, $\gamma $, $\delta $ and $\mathcal{O}$. The deformation then
reads 
\begin{equation}
\mathcal{L}[\varphi ]=\mathcal{L}_{\mathcal{C}}[\varphi ,\alpha ]+\overline{%
\lambda }_{\ell }^{I}\mathcal{O}_{\ell I}(\varphi )-\sum_{n=2}^{\infty
}(\gamma _{n\ell }^{-1})^{IJ}\delta _{n,J}~\mathcal{O}_{n\ell ,I}(\varphi ),
\label{def2}
\end{equation}
where appropriate summations over $I,J\ldots $ are understood. By
assumption, the matrix $\gamma _{\ell }^{IJ}$ should have a null eigenvector 
$\overline{\lambda }_{\ell }^{I}$. Then the operator $\overline{\lambda }%
_{\ell }^{I}\mathcal{O}_{\ell I}$ is finite and, used as the head of the
queue, also protected, so the deformation is non-trivial. Instead, the
matrices $\gamma _{n\ell }$, $n>1$, should be invertible unless $\delta
_{n}=0$.

\bigskip

Assume that $\alpha =0$ is the free-field limit of $\mathcal{C}$, and that $%
\alpha $ is defined so that the anomalous dimensions $\gamma _{n\ell }$ are
generically of order $\alpha $. In the presence of three-leg marginal
vertices in four dimensions (multiplied by a coupling $g$ such that $\alpha
=g^{2}$) some non-diagonal entries of $\gamma _{n\ell }^{IJ}$ are of order $%
g $, due to the renormalization mixing. For the time being I assume that
there are no three-leg marginal vertices. The general case is treated in
subsection 4.1.

Some $\gamma _{n\ell }$ might have vanishing low-order coefficients. Call $%
q_{n}$ the lowest non-vanishing order of $\gamma _{n\ell }$ and define $%
q=\max_{n}q_{n}$. When $\alpha $ is small, the $\lambda _{n\ell }$ behave at
worst as 
\begin{equation}
\lambda _{n\ell }\sim \frac{c_{n}\overline{\lambda }_{\ell }^{n}}{\alpha
^{q(n-1)}},  \label{behave}
\end{equation}
where $c_{n}$ are constants. This result is proved by induction. It is
certainly true for $n=1$. Assume that it is true also for $\lambda _{k\ell }$%
, $k<n$. Since $\delta _{n\ell }$ depends on the $\lambda $s of the lower
levels $k\ell ,$ $k<n$, its behavior is at worst $\delta _{n\ell }\sim
\prod_{k<n}\lambda _{k\ell }{}^{n_{k}}$, with $\sum_{k<n}kn_{k}=n$, where $%
n_{k}$ are non-negative integers. Moreover, $m\equiv \sum_{k<n}n_{k}\geq 2$,
since $\delta _{n\ell }$ is at least quadratic. Therefore 
\begin{equation}
\lambda _{n\ell }=-\frac{\delta _{n\ell }}{\gamma _{n\ell }}\lesssim \frac{%
\overline{\lambda }_{\ell }^{n}}{\alpha ^{q}}\prod_{k<n}\left( \frac{c_{k}}{%
\alpha ^{q(k-1)}}\right) ^{n_{k}}=\frac{c_{n}\overline{\lambda }_{\ell }^{n}%
}{\alpha ^{q(n-m+1)}},  \label{esti}
\end{equation}
which is at worst as singular as (\ref{behave}). Thus (\ref{behave}) is
proved for arbitrary $n$. A behavior such as (\ref{behave}) is called 
\textit{perturbatively meromorphic} of order $q$ (see Appendix B for more
details).

Using this result, if $q<\infty $ the deformed lagrangian can be expressed as

\begin{equation}
\mathcal{L}[\varphi ]\sim \mathcal{L}_{\mathcal{C}}[\varphi ,\alpha ]+\alpha
^{q}\lambda _{\ell \mathrm{eff}}\mathcal{O}_{\ell }(\varphi )+\alpha
^{q}\sum_{n=2}^{\infty }c_{n}(\alpha )\lambda _{\ell \mathrm{eff}}^{n}~%
\mathcal{O}_{n\ell }(\varphi ),  \label{def3}
\end{equation}
where $\lambda _{\ell \mathrm{eff}}=\lambda _{\ell }\alpha ^{-q}$ and the
functions $c_{n}(\alpha )$ are analytic in $\alpha $. The expansion in
powers of the energy is meaningful for energies $E$ much smaller than the
``effective Planck mass'' $M_{P\text{eff}}\equiv 1/\lambda _{\ell \mathrm{eff%
}}^{1/\ell }=M_{P}\alpha ^{q/\ell }$, where $M_{P}\equiv 1/\lambda _{\ell
}^{1/\ell }$. The $\alpha \rightarrow 0$ limit at fixed $\lambda _{\ell 
\mathrm{eff}}$ is the Gaussian fixed point, where the entire deformation
disappears. On the other hand, the $\alpha \rightarrow 0$ limit at fixed $%
\lambda _{\ell }$ is singular. The effective Planck mass, and therefore the
radius of convergence of the expansion, tend to zero when $\alpha $ tends to
zero at fixed $\lambda _{\ell }$. Thus the procedure cannot be used to
construct irrelevant deformations of free-field theories.

In \cite{pap2} it has been shown that three-dimensional quantum gravity
coupled with interacting conformal matter can be quantized as a finite
theory (see also \cite{pap1}). This is due to the special properties of
spacetime in three dimensions, because the Riemann tensor can be expressed
in terms of the Ricci tensor and the scalar curvature. The theory is unique,
and therefore predictive, because there is a unique finite protected
operator, which is precisely the Einstein term. The results of the next
sections can be used to generalize the construction of \cite{pap2} and
quantize also three-dimensional quantum gravity coupled with running matter
(see section 7 for details).

In \cite{pap3} finite chiral irrelevant deformations of superconformal field
theories have been constructed in four dimensions. Such deformations are
infinitely many, because all chiral operators are finite and protected.

\section{Quasi finiteness beyond power-counting}

\setcounter{equation}{0}

When the conformal field theory $\mathcal{C}$ does not admit finite
irrelevant operators, the finiteness equations have a trivial solution. Then
it is natural to look for more general irrelevant deformations, relaxing the
finiteness conditions in some way. A possibility is to demand that only a
subset of beta functions vanish. In general, however, since the RG\ flow is
a one-parameter flow, freezing one or some couplings freezes the entire flow
to a point. This is consistent only if such a point is a fixed point of the
flow, where all beta functions vanish. In less generic situations, when the
set of couplings can be divided into two subsets $g_{i}$, and $\lambda _{j}$%
, such that the $g$-beta functions admit a factorization $\beta
_{g}=h(g)f(g,\lambda )$, then it is meaningful to impose $\beta _{g}=0$ at
non-zero $\beta _{\lambda }$ solving $h(g)=0$. If $g_{i}=\overline{g}_{i}=%
\mathrm{constant}$ denote the solutions of $h(g)=0$, the RG flow of the
couplings $\lambda $ is non-trivial and consistently determined by the beta
functions $\beta _{\lambda }(\overline{g},\lambda )$.

Examples of this kind are the quasi-finite irrelevant deformations of
interacting conformal field theories \cite{pap3}. Again, the deformation is
made of a head and a queue. The head is the irrelevant term with the lowest
level, say $\ell $, multiplied by the irrelevant coupling $\lambda _{\ell }$%
, which is free to run. The queue runs coherently with the head and is made
of terms of levels $n\ell $, with $n$ integer, multiplied by unique
functions of $\lambda _{\ell }$ and the marginal couplings $\alpha $ of $%
\mathcal{C}$. Since the irrelevant couplings $\lambda _{n\ell }$ are
dimensionful, they can be conveniently split into an energy scale $1/\kappa $
and dimensionless ratios $r_{n}(\alpha )$: 
\begin{equation}
\lambda _{\ell }=\kappa ^{\ell },\qquad \lambda _{n\ell }=r_{n}~\lambda
_{\ell }^{n}.  \label{quasi}
\end{equation}
The structure of the beta functions $\beta _{r_{n}}$ of the dimensionless
couplings $r_{n}$ are immediately derived from (\ref{betagel}). Clearly, $%
\beta _{r_{n}}$ cannot depend on $\kappa $, for dimensional reasons.
Moreover, $\beta _{r_{n}}$ is linear in $r_{n}$ and at least quadratic in $%
r_{k}$ with $k<n$, while the beta function of $\kappa $ is proportional to $%
\kappa $: 
\begin{equation}
\beta _{r_{n}}(r,\alpha )=\gamma _{r_{n}}(\alpha )r_{n}+d_{n}(r_{<},\alpha
),\qquad \beta _{\kappa }=\frac{1}{\ell }\kappa \gamma _{\ell }(\alpha ),
\label{ta}
\end{equation}
where $d_{n}(r_{<},\alpha )$ depends only on $r_{k}$ with $k<n$ and $\gamma
_{r_{n}}(\alpha )=\gamma _{n\ell }(\alpha )-n\gamma _{\ell }(\alpha )$. Then
it is consistent to impose the \textit{quasi-finiteness }conditions 
\begin{equation}
\beta _{r_{n}}(r,\alpha )=0.  \label{quasifiniteness}
\end{equation}
The solutions $\overline{r}(\alpha )$ exists if certain invertibility
conditions, discussed below, hold. The lagrangian of the irrelevant
deformation then reads 
\begin{equation}
\mathcal{L}[\varphi ]=\mathcal{L}_{\mathcal{C}}[\varphi ,\alpha ]+\kappa
^{\ell }\mathcal{O}_{\ell }(\varphi )+\sum_{n=2}^{\infty }\overline{r}%
_{n}(\alpha )~\kappa ^{n\ell }~\mathcal{O}_{n\ell }(\varphi ).  \label{add}
\end{equation}
If $\mathcal{C}$ is finite, the deformed theory is renormalized by means of
field redefinitions and a unique renormalization constant, the one for $%
\kappa $. The $\kappa $-running is determined by the RG equations: 
\[
\frac{\mathrm{d}\kappa }{\mathrm{d}\ln \mu }=\beta _{\kappa }=\frac{\kappa }{%
\ell }\gamma _{\ell }(\alpha ),\qquad \kappa (\mu )=\kappa (\overline{\mu }%
)\left( \frac{\mu }{\overline{\mu }}\right) ^{\gamma _{\ell }(\alpha )/\ell
}. 
\]
For this reason, the theory (\ref{add}) is called ``quasi finite''. It is
natural to extend the notion of quasi finiteness to the irrelevant
deformations of every (interacting) conformal field theory $\mathcal{C}$, in
particular the fixed points of RG flows.

Finite and quasi-finite deformations differ for the $\kappa $-running and
for the existence conditions (\ref{conda}) that I now discuss. Using (\ref
{quasi}) and (\ref{ta}), the equations (\ref{quasifiniteness}) give
immediately 
\begin{equation}
r_{n}(\gamma _{n\ell }-n\gamma _{\ell })=-d_{n}(r_{<},\alpha ),
\label{equat}
\end{equation}
which can be solved iteratively in $n$ if 
\begin{equation}
\gamma _{n\ell }\neq n\gamma _{\ell }.  \label{conda}
\end{equation}

These invertibility conditions require that for every $n>1$ the anomalous
dimension of $\mathcal{O}_{n\ell }$ is not $n$ times the anomalous dimension
of the head. In the absence of symmetries or special protections, this is
generically true. In any case, again, (\ref{conda}) is a restriction on the
theory $\mathcal{C}$, so in principle it is possible to say which conformal
theories admit which quasi-finite deformations before effectively turning
the deformation on.

If some term $\mathcal{O}_{\ell ^{\prime }{}}$ of the queue violates (\ref
{conda}) its coupling $\lambda _{\ell ^{\prime }}$ cannot run coherently
with $\lambda _{\ell }$ and has to be treated as a new independent
parameter. The resulting deformation is a multiple deformation, namely a
deformation with more independent heads, of levels $\ell _{1},\cdots \ell
_{k}$, whose couplings $\lambda _{\ell _{i}}$ run independently. Formula (%
\ref{quasi}) generalizes to 
\begin{equation}
\lambda _{m}=\sum_{\{n\}}r_{m}^{n_{1}\cdots n_{k}}(\alpha )\ \lambda _{\ell
_{1}}^{n_{1}}\cdots \lambda _{\ell _{k}}^{n_{k}},\qquad
\sum_{j=1}^{k}n_{j}\ell _{j}=m,\quad n_{j}\geq 0.  \label{quasi2}
\end{equation}
Here quasi-finiteness is the condition that the dimensionless coefficients $%
r_{m}^{n_{1}\cdots n_{k}}$ have vanishing beta functions. If the violations
of the invertibility conditions are finitely many, then the deformation can
be renormalized with a finite number of independent couplings, plus field
redefinitions.

As in the previous section, assume that there are no three-leg vertices,
that $\alpha =0$ is the free-field limit of $\mathcal{C}$, and that $\alpha $
is defined so that the anomalous dimensions $\gamma _{n\ell }$ are
generically of order $\alpha $. When $\alpha $ is small, the behavior (\ref
{behave}) and formula (\ref{def3}) hold also in the case of quasi-finite
deformations. If $q<\infty $ the deformation is perturbatively meromorphic
of order $q$ in $\alpha $ and it is possible to define an effective Planck
mass $M_{P\text{eff}}$ such that the perturbative expansion in powers of the
energy is meaningful for energies much smaller than $M_{P\text{eff}}$. The
Gaussian fixed point is the $\alpha \rightarrow 0$ limit at fixed $M_{P\text{%
eff}}$.

By construction, the deformation is irrelevant with respect to the Gaussian
fixed point and the weakly coupled conformal theories $\mathcal{C}$. For $%
\alpha $ large, if the invertibility conditions (\ref{conda}) are fulfilled,
the deformation is relevant, marginal or irrelevant with respect to $%
\mathcal{C}$ if the head $\mathcal{O}_{\ell }$ is a relevant ($\gamma _{\ell
}(\alpha )<-\ell $), marginal ($\gamma _{\ell }(\alpha )+\ell =0$) or
irrelevant ($\gamma _{\ell }(\alpha )>-\ell $) operator of $\mathcal{C}$,
respectively. Indeed, rescale the overall energy $E$ of correlations
functions with respect to $1/\kappa $ and $\mu $. Because of (\ref{quasi})
and (\ref{quasifiniteness}), the insertions of $\int \mathcal{O}_{n\ell
}(\varphi )$ scale with exponents $n\ell +\beta _{n\ell }/\lambda _{n\ell
}=n(\ell +\gamma _{\ell })$.

\section{Infinite reduction of couplings}

\setcounter{equation}{0}

In this section I study the infinite reduction for irrelevant deformations
of running renormalizable theories $\mathcal{R}$. Again, I assume that $%
\mathcal{R}$ contains no relevant parameter. The formulas below are written,
for simplicity, in the case that $\mathcal{R}$ contains just one marginal
coupling $\alpha $. The generalization to more marginal couplings is treated
in section 8. As usual, divide the irrelevant deformation into levels, the
level of an operator being its canonical dimensionality in units of mass
minus the spacetime dimension $d$. For definiteness, I assume that $d$ is
four. The generalization to odd and other even dimensions is simple and left
to the reader.

Let \textrm{O}$_{\lambda }$ denote a basis of essential, local, symmetric,
scalar, canonically irrelevant operators constructed with the fields of $%
\mathcal{R}$ and their derivatives. I\ assume that each level contains a
finite number of terms. These in general mix under renormalization. To
simplify the notation, I collectively\ denote the operators of level $n$
with \textrm{O}$_{\lambda _{n}}$, $n=1,2,\cdots ,\infty $, and their
couplings with $\lambda _{n}$. When necessary, operators of the same level $%
n $ can be distinguished with a second label $I=1,\ldots N_{n}$ as shown in
formula (\ref{def1}).

Again, the beta functions of the non-renormalizable theory 
\begin{equation}
\mathcal{L}_{\mathrm{cl}}[\varphi ]=\mathcal{L}_{\mathcal{R}}[\varphi
,\alpha ]+\sum_{n}\lambda _{n}\mathcal{O}_{n}(\varphi )  \label{nonra}
\end{equation}
have the structure (\ref{betagel}) 
\begin{equation}
\beta _{\lambda _{n}}(\alpha ,\lambda )=\gamma _{n}\left( \alpha \right)
\lambda _{n}+\delta _{n}(\lambda _{m<n},\alpha ),  \label{reft}
\end{equation}
where $\delta _{n}$ depends only on $\lambda _{p}$ with $p<n$ and $\alpha $,
and is polynomial, at least quadratic, in the irrelevant couplings, while $%
\gamma _{n}(\alpha )$ is the anomalous dimension of \textrm{O}$_{\lambda
_{n}}$, calculated in the undeformed theory $\mathcal{R}$.

As usual, the irrelevant deformation is made of a head $\mathcal{O}_{\ell
}(\varphi )$, which is the irrelevant term with the lowest level, $\ell $,
and a queue. By dimensional arguments, the queue of the deformation is made
only of terms whose levels are integer multiples of $\ell $.

An \textit{infinite reduction} of couplings is a set of functions 
\begin{equation}
\lambda _{n\ell }=\lambda _{n\ell }(\lambda _{\ell },\alpha )=f_{n}(\alpha
)\lambda _{\ell }^{n},\qquad n>1,  \label{anz}
\end{equation}
subject to the conditions discussed below, such that the theory 
\begin{equation}
\mathcal{L}_{\mathrm{cl}}[\varphi ]=\mathcal{L}_{\mathcal{R}}[\varphi
,\alpha ]+\lambda _{\ell }\mathcal{O}_{\ell }(\varphi )+\sum_{n=2}^{\infty
}f_{n}(\alpha )\lambda _{\ell }^{n}\mathcal{O}_{n\ell }(\varphi )
\label{irrdef}
\end{equation}
is renormalized by means of field redefinitions plus renormalization
constants for $\alpha $ and $\lambda _{\ell }$.

The beta functions (\ref{reft}) read 
\begin{equation}
\beta _{n\ell }=\lambda _{\ell }^{n}\left[ f_{n}(\alpha )\gamma _{n\ell
}(\alpha )+\delta _{n}(f,\alpha )\right] ,  \label{gret}
\end{equation}
where $\delta _{n}(f,\alpha )$ depends polynomially, at least quadratically,
on $f_{k}$ with $k<n$, does not depend on $f_{k}$ with $k\geq n$. Formulas (%
\ref{anz}) and (\ref{gret}) hold also for $n=1$, if $f_{1}=1$ and $\delta
_{1}=0$.

Differentiating the functions (\ref{anz}) with respect to the dynamical
scale $\mu $ and using (\ref{gret}) the RG consistency equations 
\begin{equation}
\left[ \beta _{\alpha }\frac{\mathrm{d}}{\mathrm{d}\alpha }-\gamma _{n\ell
}(\alpha )+n\gamma _{\ell }(\alpha )\right] f_{n}(\alpha )=\delta
_{n}(f,\alpha )  \label{rg}
\end{equation}
are obtained.

Now I prove that (\ref{rg}) are necessary and sufficient conditions to
renormalize the theory by means of renormalization constants just for $%
\lambda _{\ell }$ and $\alpha $, plus field redefinitions. It is sufficient
to focus the attention on the logarithmic divergences, setting the
power-like divergences aside. Indeed, the power-like divergences are RG
invariant and can be unambiguously subtracted away just as they come,
without introducing new independent couplings. The logarithmic divergences
can be studied at the level of the renormalization group, because the
logarithms of the subtraction point $\mu $ are in one-to-one correspondence
with the logarithms of the cut-off $\Lambda $.

Write the bare couplings $\lambda _{n\ell }(\Lambda )$ and $\alpha (\Lambda
) $ in terms of their renormalization constants $Z_{n\ell }$ and $Z_{\alpha
} $ in the minimal subtraction scheme, 
\[
\lambda _{n\ell }(\Lambda )=\lambda _{n\ell }Z_{n\ell }(\lambda ,\alpha ,\ln
\Lambda /\mu ),\qquad \alpha (\Lambda )=\alpha Z_{\alpha }(\alpha ,\ln
\Lambda /\mu ), 
\]
$\lambda _{n\ell }$ and $\alpha $ being the renormalized couplings at the
subtraction point $\mu $. The renormalization of $\lambda _{n\ell }$ is not
necessarily multiplicative (only the product $\lambda _{n\ell }Z_{n\ell }$
is analytic in $\lambda _{n\ell }$, for $n>1$), but the compact notation $%
\lambda _{n\ell }Z_{n\ell }$ for $\lambda _{n\ell }(\Lambda )$ is convenient
for the purposes of the infinite reduction. A more explicit notation is e.g. 
$\lambda _{n\ell }(\Lambda )=\lambda _{n\ell }+\Delta _{n\ell }(\lambda
,\alpha ,\ln \Lambda /\mu )$, with $\Delta _{n\ell }$ analytic in the
couplings. The renormalization of $\lambda _{\ell }$ is obviously
multiplicative.

Now, assume that the couplings $\lambda _{n\ell }$ are not independent, but
satisfy (\ref{anz}) and (\ref{rg}). The RG\ consistency conditions (\ref{rg}%
) imply that the reduction relations have the same form at every energy
scale, in particular at $\mu $ and $\Lambda $. Consequently, 
\begin{equation}
\lambda _{n\ell }Z_{n\ell }=\lambda _{n\ell }(\Lambda )=f_{n}(\alpha
(\Lambda ))\lambda _{\ell }^{n}(\Lambda )=\lambda _{\ell }^{n}Z_{\ell
}^{n}f_{n}(\alpha Z_{\alpha }),\qquad n>1.  \label{areno}
\end{equation}
This formula shows that the couplings $\lambda _{n\ell }$, $n>1$, can be
renormalized just attaching renormalization constants to $\lambda _{\ell }$
and $\alpha $. The renormalization constants $Z_{n\ell }$, $n>1$, are not
independent. Indeed, (\ref{areno}) implies 
\[
Z_{n\ell }=Z_{\ell }^{n}\frac{f_{n}(\alpha Z_{\alpha })}{f_{n}(\alpha )}. 
\]

Despite these facts, no true reduction of couplings is achieved simply
solving the RG consistency conditions (\ref{rg}). Indeed, (\ref{rg}) are
differential equations for the unknown functions $f_{n}(\alpha )$, $n>1$.
The solutions depend on arbitrary constants $\xi $. From the point of view
of renormalization, the arbitrary constants $\xi $ are finite parameters,
namely $Z_{\xi }\equiv 1$. The equations (\ref{rg}) and the arguments
leading to (\ref{areno}) are simply a rearrangement of the renormalization
of the theory, with no true gain, because the number of renormalization
constants is reduced at the price of introducing new functions $f_{n}$. To
remove the $\xi $-ambiguities contained in the solutions of (\ref{rg}) and
achieve a true reduction of couplings, extra assumptions have to be made.
Guided by the experience of finite and quasi-finite theories, it is natural
to remove the $\xi $-arbitrariness requiring that the solution $f_{n}(\alpha
)$ be meromorphic in $\alpha $.

For the moment I\ assume that $\beta _{\alpha }^{(1)}\neq 0$ and that $%
\mathcal{R}$ contains only four-leg marginal vertices, i.e. it is the $%
\varphi ^{4}$ theory in four dimensions (but similar arguments apply if $%
\mathcal{R}$ the $\varphi ^{6}$ theory in three dimensions). The marginal
coupling $\alpha $ is defined so that the beta function and the anomalous
dimensions of $\mathcal{R}$ have expansions 
\begin{equation}
\beta _{\alpha }=\alpha ^{2}\beta _{\alpha }^{(1)}+\mathcal{O}(\alpha
^{3}),\qquad \gamma _{n}(\alpha )=\alpha \gamma _{n}^{(1)}+\mathcal{O}%
(\alpha ^{2}),  \label{tz2}
\end{equation}
etc. The models with marginal three-leg vertices (multiplied by $g$ such
that $\alpha =g^{2}$) are treated in subsection 4.1. I prove that if the
invertibility conditions 
\begin{equation}
r_{n,\ell }\equiv \frac{1}{\beta _{\alpha }^{(1)}}\left( \gamma _{n\ell
}^{(1)}-n\gamma _{\ell }^{(1)}\right) +n-1\notin \mathbb{N},\qquad n>1,
\label{conditions}
\end{equation}
are fulfilled, there exist unique meromorphic reduction relations of the
form 
\begin{equation}
\lambda _{n\ell }=f_{n}(\alpha )\lambda _{\ell }^{n}=\frac{\lambda _{\ell
}^{n}}{\alpha ^{n-1}}\sum_{k=0}^{\infty }d_{n,k}\alpha ^{k},  \label{bit}
\end{equation}
with unambiguous numerical coefficients $d_{n,k}$.

The result is proved by induction. Clearly, (\ref{bit}) is true for $n=1$.
Assume that $\lambda _{j\ell }$ with $j<n$ satisfy (\ref{bit}). By the usual
dimensional arguments $\delta _{n\ell }\sim \prod_{j<n}\lambda _{j\ell
}{}^{n_{j}}(1+\mathcal{O}(\alpha ))$, with $\sum_{j<n}jn_{j}=n$, where $%
n_{j} $ are non-negative integers, and $m\equiv \sum_{j<n}n_{j}\geq 2$,
since $\delta _{n\ell }$ is at least quadratic. Therefore for small $\alpha $
the inductive hypothesis implies 
\[
\delta _{n\ell }\sim \lambda _{\ell }^{n}\prod_{j<n}\left( \frac{1}{\alpha
^{j-1}}\right) ^{n_{j}}=\frac{\lambda _{\ell }^{n}}{\alpha ^{n-m}}\leq \frac{%
\lambda _{\ell }^{n}}{\alpha ^{n-2}}. 
\]
Now, insert the ansatz (\ref{bit}) into the RG consistency conditions (\ref
{rg}) and solve for $d_{n,k}$ recursively in $k$. If the invertibility
conditions (\ref{conditions}) hold, the solution is well defined and the
coefficients $d_{n,k}$ have unambiguous values of the form 
\begin{equation}
d_{n,k}=\frac{P_{n,k}}{\prod_{j=1}^{k+1}\left( \gamma _{n\ell
}^{(1)}-n\gamma _{\ell }^{(1)}+(n-j)\beta _{\alpha }^{(1)}\right) },
\label{dnk}
\end{equation}
where $P_{n,k}$ depends polynomially on the coefficients of the beta
function and the anomalous dimensions and on $d_{m,k}$ with $m<n$. This
proves the statement for arbitrary $n$. In general the numerator in (\ref
{dnk}) does not vanish when the denominator vanishes.

Formula (\ref{bit}) shows that the irrelevant deformation (\ref{irrdef}) is
perturbatively meromorphic of order one. Defining $\lambda _{\ell \mathrm{eff%
}}=\lambda _{\ell }/\alpha =1/M_{P\text{eff}}^{\ell }$, the behavior of the
lagrangian for small $\alpha $ is 
\begin{equation}
\mathcal{L}[\varphi ]\sim \mathcal{L}_{\mathcal{R}}[\varphi ,\alpha ]+\alpha
\lambda _{\ell \mathrm{eff}}\mathcal{O}_{\ell }(\varphi )+\alpha
\sum_{n=2}^{\infty }d_{n,0}\lambda _{\ell \mathrm{eff}}^{n}~\mathcal{O}%
_{n\ell }(\varphi ).  \label{repara}
\end{equation}
The perturbative expansion is meaningful for $\alpha \ll 1$, for energies $%
E\ll M_{P\text{eff}}$. Therefore, at fixed $\lambda _{\ell \mathrm{eff}}(\mu
)$ the irrelevant deformation disappears in the limit where the
renormalizable sector becomes free.

Clearly, the invertibility conditions (\ref{conditions}) are sufficient to
have a meaningful reduction that is perturbatively meromorphic of order one.
The conditions (\ref{conditions}) are not necessary, because in some cases $%
\delta _{n}$ might start from higher orders in $\alpha $, and $f_{n}$ be
less singular than (\ref{bit}). Formula (\ref{bit}) just describes the worst
behavior. Observe that the quantities $r_{n,\ell }$ depend only on one-loop
coefficients, yet they~determine the existence of the reduction to all
orders. Moreover, the $r_{n,\ell }$s are just rational numbers and it is not
unfrequent that they coincide with natural numbers for some $n$s. The
violations of the invertibility conditions can be cured introducing new
independent couplings (see below).

The parametrization (\ref{repara}) in terms of $\alpha $ and $\lambda _{\ell 
\mathrm{eff}}$ is non-minimal, in the sense that no irrelevant vertex is
multiplied by $\lambda _{\ell \mathrm{eff}}$, all irrelevant vertices being
multiplied by products $\alpha \lambda _{\ell \mathrm{eff}}^{n}$. An example
of minimal parametrization of the space of couplings is (\ref{irrdef}),
where the head is multiplied just by $\lambda _{\ell }$. All minimal
parametrizations are singular for $\alpha \rightarrow 0$. The reason is that
marginal and irrelevant deformations do not commute. In particular, it is
necessary to have $\alpha \neq 0$ to build the irrelevant deformation, but
the marginal interaction exists also in the absence of irrelevant
deformations.

\subsection{\textbf{Three-leg marginal vertices and renormalization mixing}}

Taking care of the renormalization mixing, in the absence of three-leg
marginal vertices the invertibility conditions become a straightforward
matrix generalization of (\ref{conditions}) \cite{pap6}. Instead, when $%
\mathcal{R}$ contains three-leg marginal vertices, multiplied by a coupling $%
g$ such that $\alpha =g^{2}$ (e.g. $\mathcal{R}$ is a gauge theory), the
effects of the renormalization mixing are non-trivial. It is convenient to
define a parity transformation, called $U$, that sends $g$ into $-g$ and
every field $\varphi $ into $-\varphi $. Clearly, $\mathcal{R}$ is $U$%
-invariant. Assigning suitable $U$-parities to the irrelevant couplings $%
\lambda _{n}$ also (\ref{nonra}) is $U$-invariant. Observe that $\delta
_{n}(\lambda _{m<n},\alpha )$ can contain non-negative powers of both $%
\alpha $ and $g$. To simplify the treatment, it is convenient to work with $%
U $-even quantities whenever possible, which can be achieved with a simple
trick. Define $\widehat{\mathcal{O}}_{n\ell }(\varphi )\equiv g^{N_{n\ell
}-2}\mathcal{O}_{n\ell }(\varphi )$ and $\widehat{\lambda }_{n\ell }$ such
that $\widehat{\lambda }_{n\ell }\widehat{\mathcal{O}}_{n\ell }(\varphi
)=\lambda _{n\ell }\mathcal{O}_{n\ell }(\varphi )$, where $N_{n\ell }$ is
the number of legs of the vertex $\mathcal{O}_{n\ell }(\varphi )$.
Generalizing (\ref{anz}) to 
\begin{equation}
\widehat{\lambda }_{n\ell }=\widehat{\lambda }_{n\ell }(\widehat{\lambda }%
_{\ell },\alpha )=\widehat{f}_{n}(\alpha )\widehat{\lambda }_{\ell
}^{n},\qquad n>1,  \label{anz2}
\end{equation}
it is clear that every $\widehat{f}_{n}$ is even. Simple diagrammatics show
that $\delta _{n\ell }(\lambda ,\alpha )=g^{N_{n\ell }}\widehat{\delta }%
_{n\ell }(\widehat{\lambda },\alpha )$, where $\widehat{\delta }_{n\ell }(%
\widehat{\lambda },\alpha )$ is polynomial in the $\widehat{\lambda }_{k\ell
}$'s, $k<n$, analytic in $\alpha $ and does not depend on the $\widehat{%
\lambda }_{k\ell }$'s with $k>n$. Indeed, let $G$ be a diagram contributing
to $\delta _{n\ell }$, with $E$ external legs, $I$ internal legs and $V$
vertices. The $g$-powers carried by the vertices are equal to $E+2I$, which
is the number of legs carried by the vertices, minus $2V$. Since $%
I-V=L-1\geq 0$, where $L$ is the number of loops, and $N_{n\ell }=E$, the
result follows. By the same argument, the anomalous dimensions $\widehat{%
\gamma }_{n\ell }$ of the operators $\widehat{\mathcal{O}}_{n\ell }(\varphi
) $ are analytic in $\alpha $ and at one loop they are of order $\alpha $.
The reduction equations for the $\widehat{f}_{n}$'s are then 
\begin{equation}
\left[ \beta _{\alpha }\frac{\mathrm{d}}{\mathrm{d}\alpha }-\widehat{\gamma }%
_{n\ell }(\alpha )+n\widehat{\gamma }_{\ell }(\alpha )\right] \widehat{f}%
_{n}(\alpha )=\alpha \widehat{\delta }_{n\ell }(\widehat{f},\alpha ).
\label{redu}
\end{equation}
If the invertibility contitions 
\begin{equation}
\frac{1}{\beta _{\alpha }^{(1)}}\left( \widehat{\gamma }_{n\ell }^{(1)}-n%
\widehat{\gamma }_{\ell }^{(1)}\right) \notin \mathbb{N},\qquad n>1,
\label{invodd}
\end{equation}
hold, the equations (\ref{redu}) admit unique solutions $\widehat{f}%
_{n}(\alpha )$ analytic in $\alpha $. In this parametrization, $\widehat{%
\lambda }_{\ell }$ coincides with $\lambda _{\ell \mathrm{eff}}$. Since each
irrelevant vertex has at least three legs, the deformation 
\begin{equation}
\mathcal{L}[\varphi ]\sim \mathcal{L}_{\mathcal{R}}[\varphi ,\alpha
]+g^{N_{\ell }-2}\lambda _{\ell \mathrm{eff}}\mathcal{O}_{\ell }(\varphi
)+\sum_{n=2}^{\infty }g^{N_{n\ell }-2}\widehat{f}_{n}(\alpha )\lambda _{\ell 
\mathrm{eff}}^{n}~\mathcal{O}_{n\ell }(\varphi )  \label{yeti}
\end{equation}
is perturbatively meromorphic of order $g$, instead of $\alpha $.

Now, consider the renormalization mixing, calculated in the undeformed
theory $\mathcal{R}$, among operators with the same dimensionality $n\ell $
in units of mass, $n\geq 1$. Distinguish the mixing operators with indices $%
I,J,\ldots $. If $\ell $ is the level of the deformation, denote the
inequivalent operators of level $\ell $ with $\widehat{\mathcal{O}}_{\ell
}^{I}$, the coefficient-matrix of their one-loop anomalous dimensions with $%
\widehat{\gamma }_{\ell }^{(1)\ IJ}$, an eigenvalue of $\widehat{\gamma }%
_{\ell }^{(1)\ IJ}$ with $\gamma _{\ell }^{(1)}$ and the corresponding
eigenvector with $d_{0}^{I}$. For simplicity, assume that $\gamma _{\ell
}^{(1)}$ is real. Below I describe how the arguments are modified when $%
\gamma _{\ell }^{(1)}$ is complex. Denote the operators of the queue of the
deformation with $\widehat{\mathcal{O}}_{n\ell }^{I}$, $n>1$, and their
couplings with $\widehat{\lambda }_{n\ell }^{I}$. In the hatted notation
introduced above the matrices of anomalous dimensions $\widehat{\gamma }%
_{n\ell }^{IJ}$ are analytic in $\alpha $ and at one loop they are of order $%
\alpha $. The hatted beta functions read 
\[
\widehat{\beta }_{n\ell }^{I}=\sum_{J}\widehat{\gamma }_{n\ell }^{IJ}(\alpha
)\widehat{\lambda }_{n\ell }^{J}+\alpha \widehat{\delta }_{n\ell }^{I}, 
\]
where $\widehat{\delta }_{n\ell }^{I}$ depends only on $\widehat{\lambda }%
_{m\ell }^{I}$ with $m<n$ and is analytic in $\alpha $. Introduce an
auxiliary coupling $\widehat{\lambda }_{\ell }$ of level $\ell $, with beta
function $\widehat{\beta }_{\lambda _{\ell }}=\gamma _{\ell }^{(1)}\alpha 
\widehat{\lambda }_{\ell }$. The beta function of $\widehat{\lambda }_{\ell
} $ can be chosen to be one-loop exact with an approprite scheme choice (any
other choice being equivalent to a redefinition $\widehat{\lambda }_{\ell
}\rightarrow h(\alpha )\widehat{\lambda }_{\ell }$, with $h(\alpha )$
analytic in $\alpha $, $h(0)=1$). The reduction relations have the form 
\[
\widehat{\lambda }_{n\ell }^{I}=\widehat{f}_{n}^{I}(\alpha )\widehat{\lambda 
}_{\ell }^{n},\qquad n\geq 1, 
\]
where $\widehat{f}_{n}^{I}(\alpha )$ are analytic in $\alpha $. If $k$ is a
natural number, it is straightforward to check that the existence conditions
are that the matrices 
\begin{equation}
\widehat{r}_{n,k,\ell }^{IJ}=\widehat{\gamma }_{n\ell }^{(1)\ IJ}-n\gamma
_{\ell }^{(1)}\delta ^{IJ}-k\beta _{\alpha }^{(1)}\delta ^{IJ},
\label{matrix}
\end{equation}
be invertible for $n>1$, $k\geq 0$ and for $n=1$, $k>0$. If the
invertibility conditions are fulfilled, the solution is uniquely determined
in terms of $d_{0}^{I}$. The head of the deformation is $\sum_{I}\widehat{%
\mathcal{O}}_{\ell }^{I}\widehat{\lambda }_{\ell }^{I}$.

The entries of the matrices $\widehat{\gamma }_{n\ell }^{(1)\ IJ}$ are
rational numbers divided by $\pi ^{d/2}$. For the purposes of the infinite
reduction, the renormalization mixing is non-trivial when the matrix $%
\widehat{\gamma }_{n\ell }^{(1)\ IJ}$ is non-triangular. In general, the
size of the non-triangular blocks of $\widehat{r}_{n,k,\ell }^{IJ}$ grows
with $n$. A renormalization mixing with these properties makes the
violations of the existence conditions much rarer, since the eigenvalues of
a non-triangular matrix with rational entries are in general irrational or
complex. Below I\ explain that any time the invertibility conditions are
violated a new coupling has to be introduced. It is reasonable to expect
that a sufficiently non-trivial renormalization mixing causes at most the
sporadic appearance of a finite number of new couplings.

Multiple-head deformations are treated as explained at the end of section 3,
see formula (\ref{quasi2}). If the eigenvalue $\gamma _{\ell }^{(1)}$ is
complex it is necessary to consider its complex conjugate $\overline{\gamma }%
_{\ell }^{(1)}$ and the corresponding eigenvector $\overline{d}_{0}^{I}$
together with $\gamma _{\ell }^{(1)}$ and $d_{0}^{I}$, and introduce the
conjugate auxiliary coupling $\widehat{\overline{\lambda }}_{\ell }$, with
beta function $\widehat{\beta }_{\overline{\lambda }_{\ell }}=\overline{%
\gamma }_{\ell }^{(1)}\alpha \widehat{\overline{\lambda }}_{\ell }$. The
reduction relations are expansions of the form (\ref{quasi2}) in powers of $%
\widehat{\lambda }_{\ell }$ and $\widehat{\overline{\lambda }}_{\ell }$, and
have to satisfy straightforward reality conditions. This gives in practice a
two-head deformation. Alternatively, it is possible to use a real two-by-two
matrix (the real Jordan canonical form) in place of $\gamma _{\ell }^{(1)}$
and then proceed as for two-head deformations, without the need of reality
conditions.

In each model, the more appropriate invertibility conditions are (\ref
{invodd}) or (\ref{conditions}) depending on the presence or absence of
marginal three-leg vertices. Perturbative meromorphy is described by (\ref
{yeti}) or (\ref{repara}), respectively. To keep the notation to a minimum,
in the rest of the paper I work in the absence of marginal three-leg
vertices, since it is straightforward to adapt the arguments to the other
case when necessary. More details can be found in \cite{halat}.

\subsection{\textbf{Dependence on the arbitrary constants and uniqueness of
the infinite reduction}}

\noindent When the invertibility conditions (\ref{conditions}) are
fulfilled, the most general solution of (\ref{rg}) is 
\begin{equation}
\frac{1}{\alpha ^{n-1}}\sum_{k=0}^{\infty }d_{n,k}\alpha ^{k}+\xi _{n}%
\overline{s}_{n}(\alpha ),  \label{mosta}
\end{equation}
where $d_{n,k}$ are given in (\ref{dnk}), $\xi _{n}$ is an arbitrary
constant and the function 
\begin{equation}
\overline{s}_{n}(\alpha )=\exp \left( \int^{\alpha }\mathrm{d}\alpha
^{\prime }\frac{\gamma _{n\ell }(\alpha ^{\prime })-n\gamma _{\ell }(\alpha
^{\prime })}{\beta _{\alpha }(\alpha ^{\prime })}\right)   \label{solb}
\end{equation}
is the solution of the homogeneous equation. For $\alpha $ small 
\[
\overline{s}_{n}(\alpha )\sim \alpha ^{Q_{n}},\qquad Q_{n}=\frac{\gamma
_{n\ell }^{(1)}-n\gamma _{\ell }^{(1)}}{\beta _{\alpha }^{(1)}}.
\]
Formula (\ref{mosta}) can be used also to study the solutions of (\ref{rg})
when the invertibility conditions are violated, with a suitable process of
limit. Three situations can take place:

$i$) If $Q_{n}$ is not integer $\overline{s}_{n}(\alpha )$ is not
meromorphic. In this case the invertibility conditions are fulfilled and
perturbative meromorphy fixes $\xi _{n}=0$, thereby selecting the reduction
uniquely.

$ii$) When $Q_{n}$ is an integer $\overline{p}\geq -n+1$ the invertibility
conditions (\ref{conditions}) are violated at order $\overline{p}+n-1$. To
study this case, use (\ref{mosta}) to approach the case $Q_{n}=\overline{p}$
from $Q_{n}=\overline{p}+\delta $, $\delta $ irrational, taking the limit $%
\delta \rightarrow 0$. The singularity $\sim 1/\delta $ in $d_{n,\overline{p}%
+n-1}$ can be removed redefining the constant $\xi _{n}$, 
\begin{equation}
\frac{1}{\alpha ^{n-1}}\frac{b_{n,\overline{p}+n-1}\alpha ^{\overline{p}+n-1}%
}{Q_{n}-\overline{p}}+\xi _{n}\alpha ^{Q_{n}}=\alpha ^{\overline{p}}\left( 
\frac{b_{n}}{\delta }+\xi _{n}\alpha ^{\delta }\right) \sim \alpha ^{%
\overline{p}}\left( -b_{n}\ln \alpha +\xi _{n}^{\prime }\right) .
\label{loga}
\end{equation}
Here $b_{n,\overline{p}+n-1}$ is a known non-singular factor. Formula (\ref
{loga}) shows that meromorphy is violated by a logarithm and no value of $%
\xi _{n}^{\prime }$ can eliminate it. The violation of meromorphy can be
reabsorbed only introducing a new independent coupling (see below), which
reabsorbs also the singularities of $d_{n,k}$ with $k>\overline{p}+n-1$. The
difference between this case and case $i$) is that here the introduction of
the new coupling is compulsory, while in case $i$) the violation of
meromorphy can be removed with an appropriate choice of the arbitrary
constant $\xi _{n}$.

$iii$) When $Q_{n}$ is an integer $<-n+1$ the invertibility conditions (\ref
{conditions}) are fulfilled and the solution (\ref{mosta}) is meromorphic
for arbitrary $\xi _{n}$. However, the order of perturbative meromorphy
increases.

Now I study cases $ii$) and $iii$) in more detail.

\subsection{\textbf{Case }$ii$)\textbf{. Violations of the invertibility
conditions and introduction of new parameters at higher orders}}

\noindent Suppose that $r_{\overline{n},\ell }$ is a natural number $%
\overline{k}$ for some $\overline{n}$ or that some matrix $r_{n,k,\ell
}^{IJ} $ has a null eigenvector. Then the reduction fails at the $\overline{k%
}$th order, unless a new independent parameter $\overline{\lambda }_{%
\overline{n}\ell }$ is introduced at that order in front of $\mathcal{O}_{%
\overline{n}\ell }$. Write 
\begin{equation}
\lambda _{\overline{n}\ell }=\frac{1}{\alpha ^{\overline{n}-1}}\left[
\lambda _{\ell }^{\overline{n}}\sum_{j=0}^{\overline{k}-1}d_{\overline{n}%
,j}\alpha ^{j}+\alpha ^{\overline{k}}\overline{\lambda }_{\overline{n}\ell
}\right] ,  \label{tre}
\end{equation}
where $d_{\overline{n},j}$, $j<\overline{k}$ are calculated as above. The
new parameter $\overline{\lambda }_{\overline{n}\ell }$ hides the logarithm
of (\ref{loga}). Its beta function has the form 
\[
\overline{\beta }_{\overline{n}\ell }=\overline{\gamma }_{\overline{n}\ell
}\left( \alpha \right) \overline{\lambda }_{\overline{n}\ell }+\overline{%
\delta }_{\overline{n}\ell }(\lambda _{m<\overline{n}},\alpha ),\qquad 
\overline{\gamma }_{\overline{n}\ell }\left( \alpha \right) =\overline{n}%
\gamma _{\ell }^{(1)}\alpha +\mathcal{O}(\alpha ^{2}),\qquad \overline{%
\delta }_{\overline{n}\ell }=\lambda _{\ell }^{\overline{n}}\mathcal{O}%
(\alpha ), 
\]
The one-loop coefficient of $\overline{\gamma }_{\overline{n}\ell }$ is
derived from the equality $r_{\overline{n},\ell }=\overline{k}$.

The introduction of $\overline{\lambda }_{\overline{n}\ell }$ affects also
the reduction relations for $n>\overline{n}$. Observing that the $\overline{%
\lambda }_{\overline{n}\ell }$ contributes only from order $\overline{k\text{%
,}}$ the modified reduction relations for $n>\overline{n}$ read 
\begin{equation}
\lambda _{n\ell }=\frac{1}{\alpha ^{n-1}}\sum_{q=0}^{[n/\overline{n}]}\alpha
^{\overline{k}q}a_{n\ell }^{(q)}(\alpha )\lambda _{\ell }^{n-\overline{n}q}%
\overline{\lambda }_{\overline{n}\ell }^{q},\qquad n>\overline{n},
\label{modi}
\end{equation}
where $[$ $]$ denotes the integral part and the coefficients $a_{n\ell
}^{(q)}$ are power series in $\alpha $. Inserting (\ref{modi}) in (\ref{rg})
the coefficients $a_{n\ell }^{(q)}$ are worked out iteratively from $q=[n/%
\overline{n}]$ to $q=0$, term-by-term in the $\alpha $-expansion. The
existence conditions for $a_{nm}^{(q)}$ are 
\begin{equation}
r_{n,\ell ,q}=r_{n,\ell }-\overline{k}q\notin \mathbb{N}  \label{klq}
\end{equation}
and do not add further restrictions, because they are already contained in (%
\ref{conditions}).

When a further invertibility condition (\ref{conditions}), $n>\overline{n}$,
is violated, the story repeats. A new parameter $\overline{\lambda }_{n\ell
} $ is introduced at order $\alpha ^{r_{n,\ell }}$. If several conditions (%
\ref{klq}), for different values of $q$, are violated at the same time, all
singular monomials of (\ref{modi}) are reabsorbed into the same new
parameter $\overline{\lambda }_{n\ell }$. Observe that the reduction itself
guides the introduction of the new parameters.

Due to (\ref{tre}) and (\ref{modi}), after the introduction of the new
parameters $\overline{\lambda }_{\overline{n}\ell }$ the deformation is
still perturbatively meromorphic of order one: it is sufficient to define $%
\overline{\lambda }_{\overline{n}\ell \text{eff}}=\alpha ^{-\overline{n}}%
\overline{\lambda }_{\overline{n}\ell }$ and take $\alpha $ small at fixed $%
M_{P\text{eff}}$ anf $\overline{\lambda }_{\overline{n}\ell \text{eff}}$.

\bigskip

Violations of the invertibility conditions (\ref{conditions}) or (\ref
{matrix}), although unfrequent, can occur. In some models the
renormalization mixing can be sufficiently non-trivial to keep the
violations to a finite number. Then the final theory is predictive, in the
sense that it is renormalized with a finite number of independent couplings
and renormalization constants, plus field redefinitions. It is possible to
have a form of predictivity also when the violations of the invertibility
conditions are infinitely many. Indeed, because of formula (\ref{conditions}%
), it is reasonable to expect that in general the quantity $r_{n,\ell }$
grows with $n$. This ensures that the new parameters $\overline{\lambda }_{%
\overline{n}\ell }$ are sporadically introduced at increasingly high orders
in $\alpha $. Even if the final theory contains infinitely independent
couplings, a finite subset of them is sufficient to make high-order
predictions. In ref. \cite{pap6} I have given models that illustrate these
facts. Consider the scalar theory 
\begin{equation}
\mathcal{L}=\frac{1}{2}(\partial _{\mu }\varphi )^{2}+V(\varphi ,\partial
\varphi ,\partial ^{2}\varphi ,\ldots )=\frac{1}{2}(\partial \varphi
)^{2}+\sum_{n=0}^{\infty }\!\!\!\!\!\!\left. \phantom{{a\over a}}\right.
^{\prime }V_{n}(\varphi )[\partial ^{2n}\varphi ],  \label{usta}
\end{equation}
where $[\partial ^{2n}\varphi ]$ is a compact notation to denote $2n$
variously distributed derivatives of the field $\varphi $, contracted in all
possible ways, and the primed sum runs over a basis of terms that are
inequivalent with respect to field redefinitions and additions of total
derivatives.

Assume that the head of the irrelevant deformation is the operator $\varphi
^{\ell +4}$. At one and two loops the potential $V_{0}(\varphi )$ does not
mix with derivative terms. This ensures that the invertibility conditions
for the monomials $\varphi ^{n\ell +4}$ have the form $r_{n,\ell }\notin %
\mathbb{N}$ with $r_{n,\ell }$ given by (\ref{conditions}). Inserting the
one-loop values of the anomalous dimensions and beta functions, the
invertibility conditions are 
\begin{equation}
r_{n,\ell }=\frac{1}{6}(n-1)(n\ell ^{2}-6)\notin \mathbb{N}  \label{inve}
\end{equation}
for $n>1$. The condition is violated in infinitely many cases. When $r_{%
\overline{n},\ell }=\overline{k}(\overline{n})\in \mathbb{N}$ for some $%
\overline{n}$, a new parameter is introduced at order $\overline{k}(%
\overline{n})$. For example, for $\ell =2$, which is the theory $\varphi
^{4}+\varphi ^{6}$, the first integer values of $r_{n,2}$ are 2,5,15,22,40$%
\ldots $, so the first new parameter appears at two loops. The terms with $%
n>0$ in (\ref{usta}) provide other invertibility conditions, similar to (\ref
{inve}), and from a certain point onwards the renormalization mixing becomes
non-trivial. So, formula (\ref{inve}) is sufficient to estimate the growth
of the number of parameters and ensures that it is possible to make
calculations up to forty loops using about ten independent couplings.

\subsection{\textbf{Case }$iii$)\textbf{. Properties of the effective Planck
mass}}

\noindent If the invertibility conditions are fulfilled, but $r_{\overline{n}%
,\ell }=-\overline{r}$ is a negative integer for some $\overline{n}$ then (%
\ref{mosta}) and (\ref{solb}) show that the solution $f_{\overline{n}%
}(\alpha )$ admits an arbitrary parameter $\overline{d}$ multiplying a more
singular meromorphic expansion 
\begin{equation}
f_{\overline{n}}(\alpha )=\frac{\overline{d}}{\alpha ^{\overline{n}-1+%
\overline{r}}}\sum_{k=0}^{\infty }c_{\overline{n},k}\alpha ^{k}+\frac{1}{%
\alpha ^{\overline{n}-1}}\sum_{k=0}^{\infty }d_{\overline{n},k}\alpha ^{k}.
\label{more}
\end{equation}
The coefficients $c_{\overline{n},k}$ and $d_{\overline{n},k}$ are uniquely
determined, with $c_{\overline{n},0}=1$. I assume that for $n<\overline{n}$
the functions $f_{n}$ behave as in (\ref{bit}). By Theorem B3 of Appendix B,
if some more restrictive invertibility conditions are fulfilled for $n>%
\overline{n}$ (see formula (\ref{rptimo}) below), the behavior of $%
f_{n}(\alpha )$ for arbitrary $n$ is 
\[
f_{n}(\alpha )\sim \frac{1}{\alpha ^{n-1+\overline{r}[n/\overline{n}]}} 
\]
and the irrelevant deformation with heads $\lambda _{\ell }$ and $\widehat{%
\lambda }_{\overline{n}\ell }\equiv \overline{d}\lambda _{\ell }^{\overline{n%
}}$ is perturbatively meromorphic of order 
\begin{equation}
\overline{q}=1+\left[ \frac{\overline{r}}{\overline{n}-1}\right] _{+},
\label{orde}
\end{equation}
$[x]_{+}$ denoting the minimum integer $\geq x$. The effective Planck mass
is defined by $\lambda _{\ell \mathrm{eff}}=\lambda _{\ell }\alpha ^{-%
\overline{q}}=1/M_{P\text{eff}}^{\ell }$ and $\widehat{\lambda }_{\overline{n%
}\ell \text{eff}}=\alpha ^{-\overline{n}\overline{q}}\widehat{\lambda }_{%
\overline{n}\ell }=\overline{d}/M_{P\text{eff}}^{\overline{n}\ell }$: for $%
\alpha \sim 0$%
\begin{equation}
\mathcal{L}[\varphi ]\sim \mathcal{L}_{\mathcal{R}}[\varphi ,\alpha ]+\alpha
^{\overline{q}}\lambda _{\ell \text{eff}}\mathcal{O}_{\ell }(\varphi
)+\alpha ^{\overline{q}}\sum_{n=2}^{\infty }a_{n}\alpha ^{p_{n}}\lambda
_{\ell \text{eff}}^{n}~\mathcal{O}_{n\ell }(\varphi ),  \label{behaviot}
\end{equation}
where $a_{n}$ are factors that depend also on the arbitrary parameter $%
\overline{d}$ and $p_{n}$ are non-negative integers. From (\ref{behaviot})
it follows that sufficient invertibility conditions for $n>\overline{n}$ are 
\begin{equation}
r_{n,\ell }^{\prime }\equiv \frac{1}{\beta _{\alpha }^{(1)}}\left( \gamma
_{n\ell }^{(1)}-n\gamma _{\ell }^{(1)}\right) +\overline{q}(n-1)\notin %
\mathbb{N},\qquad n>\overline{n}.  \label{rptimo}
\end{equation}

If some other $r_{n,\ell }^{\prime }$, $n>\overline{n}$, is an integer, the
procedures described so far can be iterated straightforwardly.

In the case just studied the meromorphic reduction admits arbitrary finite
parameters $\overline{d}$. When a new coupling of type $\widehat{\lambda }_{%
\overline{n}\ell }$ is introduced, the order $\overline{q}$ of perturbative
meromorphy increases. The effective Planck mass $M_{P\text{eff}}=M_{P}\alpha
^{\overline{q}/\ell }$ becomes smaller for $\alpha \rightarrow 0$ at fixed $%
M_{P}$. Equivalently, the irrelevant interaction becomes weaker for $\alpha
\rightarrow 0$ at fixed $M_{P\text{eff}}$, as shown by formula (\ref{def3}).
Since the expansion in powers of the energy is meaningful for $E\ll M_{P%
\text{eff}}$, a smaller effective Planck mass means a more restricted
perturbative domain. It is meaningful to require that the perturbative
domain be maximal, in which case all extra finite parameters $\overline{d}$
have to be switched off.

\subsection{$\mathcal{R}$\textbf{-beta function with some vanishing
coefficients}}

\noindent So far I have assumed $\beta _{\alpha }^{(1)}\neq 0$. When $\beta
_{\alpha }^{(1)}=0$ the invertibility conditions (\ref{conditions}) for the
existence of the coefficients $d_{n,k}$ in (\ref{dnk}) simplify and become
just one for every $n$, namely 
\begin{equation}
\gamma _{n\ell }^{(1)}\neq n\gamma _{\ell }^{(1)}.  \label{oki}
\end{equation}
This situation is an interesting generalization of the finite and
quasi-finite theories of sections 2 and 3: when $\beta _{\alpha }\neq 0$,
but $\beta _{\alpha }^{(1)}=0$ the conditions (\ref{conda}) are replaced by
their one-loop counterparts (\ref{oki}). Here $\overline{s}_{n}(\alpha )$
has an essential singularity, so perturbative meromorphy implies $\xi _{n}=0$
for every $n$.

When $\beta _{\alpha }^{(1)}=0$ and (\ref{oki}) are fulfilled for $n\neq 
\overline{n}$, but violated for $n=\overline{n}$, the invertibility
conditions for $f_{\overline{n}}(\alpha )$ are 
\begin{equation}
r_{\overline{n},\ell }^{(2)}=Q_{\overline{n}}^{(2)}+\overline{n}\equiv \frac{%
1}{\beta _{\alpha }^{(2)}}\left( \gamma _{\overline{n}\ell }^{(2)}-\overline{%
n}\gamma _{\ell }^{(2)}\right) +\overline{n}\notin \mathbb{N},  \label{r2}
\end{equation}
assuming $\beta _{\alpha }^{(2)}\neq 0$. Then $f_{\overline{n}}(\alpha )$
has a more singular expansion 
\[
f_{\overline{n}}(\alpha )=\frac{1}{\alpha ^{\overline{n}}}\sum_{k=0}^{\infty
}d_{\overline{n},k-1}\alpha ^{k}. 
\]
The invertibility conditions for $n>\overline{n}$ are not modified. Applying
theorem B3 of Appendix B with $q=1$ and $\overline{r}=1$ it follows that
when the unmodified existence conditions for $n>\overline{n}$ are fulfilled,
(\ref{bit}) is replaced by the more singular expansion 
\begin{equation}
f_{n}(\alpha )=\frac{1}{\alpha ^{n-1+[n/\overline{n}]}}\sum_{k=0}^{\infty
}d_{n,k-[n/\overline{n}]}\alpha ^{k}.  \label{sing}
\end{equation}
The coefficients $d_{n,k-[n/\overline{n}]}$ are uniquely determined and the
deformation is perturbatively meromorphic of order two, the effective Planck
mass being $\lambda _{\ell \mathrm{eff}}=\lambda _{\ell }/\alpha ^{2}=1/M_{P%
\text{eff}}^{\ell }$.

Here $\overline{s}_{\overline{n}}(\alpha )\sim \alpha ^{Q_{\overline{n}%
}^{(2)}}$, so when $r_{\overline{n},\ell }^{(2)}$ is not ingeter
perturbative meromorphy implies $\xi _{\overline{n}}=0$, when $r_{\overline{n%
},\ell }^{(2)}=\overline{k}\in \mathbb{N}$ it is compulsory to introduce a
new independent coupling at order $\overline{k}$ and when $r_{\overline{n}%
,\ell }^{(2)}=-\overline{r}$ is a negative integer it is possible to add an
arbitrary parameter at the price of increasing the order of perturbative
meromorphy.

If $\beta _{\alpha }^{(1)}=0$, $\gamma _{\overline{n}\ell }^{(1)}=\overline{n%
}\gamma _{\ell }^{(1)}$ and $\beta _{\alpha }^{(2)}=0$ then the existence
conditions become 
\[
\gamma _{\overline{n}\ell }^{(2)}\neq \overline{n}\gamma _{\ell }^{(2)}, 
\]
and so on. If $\beta _{\alpha }\equiv 0$, the procedure just described can
be iterated until a $k$ is found such that $\gamma _{\overline{n}\ell
}^{(k)}-\overline{n}\gamma _{\ell }^{(k)}\neq 0$. Only when $\gamma _{%
\overline{n}\ell }^{(k)}-\overline{n}\gamma _{\ell }^{(k)}=0$ for every $k$
it is necessary to introduce a new parameter. If $k_{n}$ denotes the minimum
integer such that $\gamma _{n\ell }^{(k_{n})}-n\gamma _{\ell }^{(k_{n})}\neq
0$ and $\overline{k}$ denotes the maximum $k_{n}$, then the deformation is
perturbatively meromorphic of order $\overline{k}$. The properties of finite
and quasi-finite irrelevant deformations, which are precisely the case $%
\beta _{\alpha }\equiv 0$, are thus recovered.

\bigskip

In conclusion, the infinite reduction works in most models and its main
properties are very general, although the details depend on the particular
model. The existence conditions have the form (\ref{conditions}), (\ref
{invodd}) or require the invertibility of matrices such as (\ref{matrix}),
whose entries are rational numbers divided by $\pi ^{d/2}$, in even
dimensions $d$. Generically, such conditions are violated only in sporadic
cases. A sufficiently non-trivial renormalization mixing can make the
infinite reduction work with a finite number of independent couplings. If
this is not the case, the number of independent couplings can grow together
with the order of the perturbative expansion, and the final theory can
contain infinitely many independent couplings, but in general the growth is
slow, in the sense that a reasonably small number of couplings are
sufficient to make calculations up to very high orders. Thus, the
infinite-reduction prescription enhances the predictive power considerably
with respect to the usual formulation of non-renormalizable theories, where
infinitely many independent couplings are present already at the tree level.

It is worth to mention that in the realm of renormalizable theories, the
existence conditions for Zimmermann's reduction of couplings include the
requirement that a certain discriminant be non-negative. A review and
details are contained in Appendix A, see formula (\ref{co1}). Several
renormalizable theories are excluded by this restriction. In the absence of
relevant parameters, the infinite reduction does not include constraints of
this type. That is why the infinite reduction works in most models.

Finally, observe that the quantities that determine the invertibility
conditions and the behavior of the solution for $\alpha $ small (i.e. $Q_{n}$%
, $\beta _{\alpha }^{(1)}$, $\gamma _{n}^{(1)}$, etc.) are
scheme-independent. This proves that the infinite reduction is
scheme-independent.

In odd dimensions, the main modification of the results derived above is
that the one-loop coefficients of the beta functions and anomalous
dimensions that appear in the invertibility conditions (\ref{conditions})
are replaced by two-loop coefficients. Indeed, diagrams with an odd number
of loops have no logarithmic divergences in odd dimensions. The other
modifications follow straightforwardly from the arguments.

\section{Interpretation of the infinite reduction}

\setcounter{equation}{0}

In this section I give an interpretation of the results derived so far, to
better clarify the meaning of the infinite reduction.

Formulated in the ordinary way, a non-renormalizable theory contains
infinitely many independent couplings, one for each essential, local,
symmetric, scalar irrelevant operator constructed with the fields and their
derivatives. In such a situation all non-renormalizable interactions are
turned on and mixed, and the theory is predictive only as an effective field
theory. The first step towards the construction of fundamental
non-renormalizable theories is to ``diagonalize'' the non-renormalizable
interactions, for example relating the irrelevant terms in a self-consistent
and scheme-independent way. Renormalization-group invariance leads to
equations (\ref{rg}), which do relate the irrelevant couplings to one
another, but are not sufficient, by themselves, to reduce the number of
couplings. Indeed, (\ref{rg}) are ingeneral differential equations, so their
solutions contain one free independent parameter $\xi $ for each coupling
that is ``reduced''. Lucky situations are those in which the equations (\ref
{rg}) are actually algebraic, which happens when the renormalizable
subsector $\mathcal{R}$ is a conformal field theory. Moreover, if $\mathcal{R%
}$ does not contain relevant parameters, the beta functions of the
irrelevant sector are linear in their own couplings and the solution
generically exists, is unique and can be worked out iteratively.

When $\mathcal{R}$ is interacting and running, the arbitrary constants $\xi $
generically multiply non-meromorphic functions of the marginal couplings $%
\alpha $. Interactions that are not perturbatively meromorphic relatively to
one another are, in some sense, ``incommensurable''. Any attempt to merge
them produces violations of perturbative meromorphy that can be used to
unmerge them back unambiguously. In practice, perturbative meromorphy
classifies the fundamental non-renormalizable interactions and can be used
to truly reduce the number of couplings. By means of renormalization and
(relative) perturbative meromorphy, quantum field theory intrinsecally
``knows'' which interactions are which. The scheme-independence of the
infinite reduction ensures that two observers that independently apply the
reduction prescription arrive at the same conclusions.

These facts uncover the intrinsic nature of fundamental interactions. A
local monomial $H(\varphi )$ in the fields and their derivatives does not
provide a good description of a fundamental interaction $\Im $, because it
is in general unstable under renormalization. The interaction can be
stabilized under renormalization when the ``head'' $H(\varphi )$ is followed
by a queue $Q$ that runs coherently with it: 
\begin{equation}
\Im (\alpha ,\lambda ,\varphi )=\lambda H(\varphi )+\sum_{n}\lambda
_{n}(\alpha ,\lambda )Q_{n}(\varphi ).  \label{basis}
\end{equation}
The queue is the sum of (generically infinitely many) local monomials $%
Q_{n}(\varphi )$ in the fields and their derivatives, multiplied by
unambiguous functions $\lambda _{n}(\alpha ,\lambda )$ of $\lambda $ and $%
\alpha $ , that can be worked out recursively in $n$. When the invertibility
conditions studied in the previous section are fulfilled, the queue is
uniquely determined by perturbative meromorphy in $\alpha $. In other cases
new couplings $\overline{\lambda }$ are sporadically introduced along the
way, guided by the reduction mechanism itself. The theory obtained deforming 
$\mathcal{R}$ with $\Im (\alpha ,\lambda ,\varphi )$ is renormalized by
field redefinitions and renormalization constants for $\alpha $, $\lambda $
and eventually $\overline{\lambda }$.

The basis (\ref{basis}) ``diagonalizes'' the non-renormalizable
interactions, and defines, for example, \textit{the} Pauli deformation, that
is to say the interaction whose head is the Pauli term, \textit{the}
four-fermion deformation (the interaction whose head is a four-fermion
vertex), \textit{the} Majorana-mass deformation, which is useful for physics
beyond the Standard Model,\ the\textit{\ }combination of some of them, and
so on. Hopefully it will soon be possible to define the ``Newton
deformation'', which encodes quantum gravity.

At the quantum level there exists one special basis (\ref{basis}) for the
fundamental interactions, while classically all basis are equally good.
Nevertheless, this is not a selection of theories, in general, because the
non-renormalizable interactions $\Im $ are still infinitely many. The
infinite reduction does not say which interactions are switched off and
which ones are switched on in nature. In special situations the infinite
reduction can also work as a selection, as it happens, for example, in
three-dimensional quantum gravity coupled with matter (see sections 2 and
7). In more general situations there remains to find a physical criterion to
select the right irrelevant deformation, or explain why no irrelevant
deformation (that is to say the undeformed renormalizable theory $\mathcal{R}
$) is better than any irrelevant deformation. Thus the infinite reduction
is, in general, a classification of the non-renormalizable interactions, but
not a selection. It is also the basic tool to address the selection issue
and makes us hope that a better understanding of the problem of quantum
gravity in four dimensions can be achieved also.

\section{Infinite reduction of couplings around interacting fixed points}

\setcounter{equation}{0}

In the previous sections I have shown that a criterion for the infinite
reduction is perturbative meromorphy around the free-field limit. In this
section I\ study the infinite reduction in the neighborhood of an
interacting fixed point. I consider theories whose renormalizable subsector $%
\mathcal{R}$ contains a single marginal coupling $\alpha $ and interpolates
between UV\ and IR\ fixed points. For simplicity, I\ assume that one fixed
point is free and the other one is interacting, since this is the more
familiar situation and generalizations are straightforward. I show that
another criterion for the infinite reduction is analyticity around the
interacting fixed point. In general, the invertibility conditions are less
restrictive than the one found in section 4 and the number of independent
couplings of the final theory remains finite. Moreover, perturbative
meromorphy around the free fixed point and analyticity around the
interacting fixed point do not hold contemporarily, but only one at a time.

It is convenient to parametrize the beta function of $\alpha $ in the form 
\begin{equation}
\beta _{\alpha }=\alpha ^{2}\left( \alpha _{*}-\alpha \right) B(\alpha ),
\label{beta4}
\end{equation}
where $B(\alpha )$ is non-vanishing and analytic throughout the RG flow,
with $B(0)=\beta _{\alpha }^{(1)}/\alpha _{*}$ and $B(\alpha _{*})=-\beta
_{*}^{\prime }/\alpha _{*}^{2}$, $\beta _{\alpha }^{(1)}$ being the one-loop
coefficient and $\beta _{*}^{\prime }$ being the slope of the beta function
at the interacting fixed point. For definiteness, I\ assume $\alpha ,\alpha
_{*}\geq 0$ and $\beta _{\alpha }^{(1)},\beta _{*}^{\prime }\neq 0$ and that
the anomalous dimensions $\gamma _{n}(\alpha )$ are regular and finite
throughout the RG\ flow. When $\alpha $ is small, the anomalous dimensions
are, generically, of order $\alpha $, $\gamma _{n}(\alpha )=\gamma
_{n}^{(1)}\alpha +\mathcal{O}(\alpha ^{2})$. Around the interacting fixed
point, instead, they tend to constant values, $\gamma _{n}(\alpha )=\gamma
_{n}^{*}+\mathcal{O}(\alpha _{*}-\alpha )$.

Consider an irrelevant deformation (\ref{irrdef}), with reduction relations (%
\ref{anz}). Once the reduction functions $f_{k}(\alpha )=\overline{f}%
_{k}(\alpha )$ are known for $k<n$, $\delta _{n}$ is a known function of $%
\alpha $. Write $\overline{\delta }_{n}(\alpha )=$ $\delta _{n}(\overline{f}%
_{k<n}(\alpha ),\alpha )$. The solution of the RG consistency conditions (%
\ref{rg}) for $f_{n}$ reads 
\begin{equation}
f_{n}(\alpha ,\xi )=\int_{c_{n}}^{\alpha }\mathrm{d}\alpha ^{\prime }\frac{%
\overline{\delta }_{n}(\alpha ^{\prime })~s_{n}(\alpha ,\alpha ^{\prime })}{%
\beta _{\alpha }(\alpha ^{\prime })},\qquad s_{n}(\alpha ,\alpha ^{\prime })=%
\frac{\overline{s}_{n}(\alpha )}{\overline{s}_{n}(\alpha ^{\prime })},
\label{sola}
\end{equation}
where $\overline{s}_{n}(\alpha )$ is defined in (\ref{solb}) and $c_{n}$ is
the arbitrary integration constant, related in a simple way with the
constants $\xi _{n}$ used in section 4. Studying formula (\ref{solb})\
around the fixed points it is immediately found that 
\begin{equation}
\overline{s}_{n}(\alpha )=\alpha ^{Q_{n}}(\alpha _{*}-\alpha
)^{Q_{n}^{*}}U(\alpha ),  \label{solvb}
\end{equation}
where 
\[
Q_{n}=\frac{\gamma _{n\ell }^{(1)}-n\gamma _{\ell }^{(1)}}{\beta _{\alpha
}^{(1)}},\qquad Q_{n}^{*}=\frac{\gamma _{n\ell }^{*}-n\gamma _{\ell }^{*}}{%
\beta _{*}^{\prime }}, 
\]
and $U(\alpha )$ is non-vanishing and analytic throughout the RG\ flow.

Now I\ prove that if the invertibility conditions 
\begin{equation}
Q_{n}^{*}\notin \mathbb{N},\qquad n>1,  \label{invf}
\end{equation}
hold, the infinite reduction is uniquely determined by analyticity around
the interacting fixed point. Assume, by induction, that the functions $%
\overline{f}_{k}(\alpha )$ with $k<n$ are unique and analytic for $\alpha
\sim \alpha _{*}$. It if sufficient to show\ that there exists a unique
choice of $c_{n}$ such that also (\ref{sola}) is analytic for $\alpha \sim
\alpha _{*}$.

The inductive assumption ensures that $\delta _{n}(\alpha )=\delta _{n}^{*}+%
\mathcal{O}(\alpha _{*}-\alpha )$ around the interacting fixed point, where $%
\delta _{n}^{*}$ is a numerical factor. Using (\ref{beta4}) and (\ref{solvb}%
) write (\ref{sola}) as 
\begin{equation}
f_{n}(\alpha )=\overline{s}_{n}(\alpha )\int_{c_{n}}^{\alpha }\mathrm{d}%
\alpha ^{\prime }\frac{\delta _{n}(\alpha ^{\prime })U^{-1}(\alpha ^{\prime
})B^{-1}(\alpha ^{\prime })}{(\alpha ^{\prime })^{2+Q_{n}}(\alpha
_{*}-\alpha ^{\prime })^{1+Q_{n}^{*}}}.  \label{custo}
\end{equation}
The properties collected so far ensure that there exists an expansion 
\[
\alpha ^{-2-Q_{n}}\delta _{n}(\alpha )U^{-1}(\alpha )B^{-1}(\alpha
)=\sum_{k=0}^{\infty }a_{n,k}(\alpha _{*}-\alpha )^{k}. 
\]
Integrating (\ref{custo}) term-by-term it is immediate to prove that the
most general solution (\ref{sola}) has the form 
\begin{equation}
f_{n}(\alpha ,\xi )=\overline{f}_{n}(\alpha )+\xi _{n}\overline{s}%
_{n}(\alpha ),  \label{solv}
\end{equation}
where 
\[
\overline{f}_{n}(\alpha )=\alpha ^{Q_{n}}U(\alpha )\sum_{k=0}^{\infty }\frac{%
a_{n,k}(\alpha _{*}-\alpha )^{k}}{Q_{n}^{*}-k} 
\]
and $\xi _{n}$ is a constant factor related with $c_{n}$.

If (\ref{invf}) hold, the function $\overline{f}_{n}(\alpha )$ is
meaningful, and analytic, around the interacting fixed point. Instead, $%
\overline{s}_{n}(\alpha )$ is not analytic for $\alpha \sim \alpha _{*}$.
Thus analyticity selects $\xi _{n}=0$ and uniquely determines the infinite
reduction, which reads 
\begin{equation}
\mathcal{L}[\varphi ]=\mathcal{L}_{\mathcal{R}}[\varphi ,\alpha ]+\lambda
_{\ell }\mathcal{O}_{\ell }(\varphi )+\sum_{n=2}^{\infty }\lambda _{\ell
}^{n}\overline{f}_{n}(\alpha )\mathcal{O}_{n\ell }(\varphi ).  \label{defar}
\end{equation}

At the interacting fixed point 
\begin{equation}
\overline{f}_{n}(\alpha )=-\frac{\delta _{n}^{*}}{\beta _{*}^{\prime
}Q_{n}^{*}}+\mathcal{O}(\alpha _{*}-\alpha )  \label{beha}
\end{equation}
and the deformed theory (\ref{defar}) tends to the quasi-finite theory 
\begin{equation}
\mathcal{L}_{\text{quasi-finite}}[\varphi ]=\mathcal{L}_{\mathcal{R}%
}[\varphi ,\alpha _{*}]+\lambda _{\ell }\mathcal{O}_{\ell }(\varphi
)-\sum_{n=2}^{\infty }\frac{\delta _{n}^{*}}{\gamma _{n}^{*}-n\gamma _{1}^{*}%
}~\lambda _{\ell }^{n}~\mathcal{O}_{n\ell }(\varphi ),  \label{defar2}
\end{equation}
\medskip whose irrelevant couplings solve the algebraic quasi-finiteness
equations 
\[
f_{n}(\alpha ^{*})~\left( \gamma _{n}(\alpha ^{*})-n\gamma _{1}(\alpha
^{*})\right) +\delta _{n}(f(\alpha ^{*}),\alpha ^{*})=0, 
\]
obtained setting $\beta _{\alpha }=0$ in (\ref{rg}). The existence
conditions (\ref{invf}) collapse to just $\gamma _{n\ell }^{*}\neq n\gamma
_{\ell }^{*}$, i.e. (\ref{conda}). Thus the irrelevant deformations of
running theories selected by analyticity around the interacting fixed point
are consistent with the finite and quasi-finite irrelevant deformations of
interacting conformal field theories of sections 2 and 3.

Observe that the quantities that determine the invertibility conditions and
the behavior of the solution around the fixed points (i.e. $Q_{n}^{*}$, $%
\delta _{n}^{*}$, $\beta _{*}^{\prime }$, $\gamma _{n}^{*}$, etc.) are
scheme-independent, so the infinite reduction around the interacting fixed
point is scheme-independent.

\bigskip

The invertibility conditions (\ref{invf}) are less restrictive than the
invertibility conditions (\ref{conditions}) for the existence of the
infinite reduction around the free fixed point, because the numbers $%
Q_{n}^{*}$ are not rational, in general. It is unlikely that (\ref{invf})
are violated for infinitely many $n$s, so in most models the analytic
reduction around the interacting fixed point produces a theory whose
divergences can be renormalized with a finite number of independent
couplings, plus field redefinitions.

If (\ref{invf}) is violated for some $\overline{n}$, $Q_{\overline{n}}^{*}=%
\overline{k}\in \mathbb{N}$, a new coupling $\overline{\lambda }_{\overline{n%
}\ell }$ has to be introduced at order $\overline{k}$ in $\alpha -\alpha
^{*} $, with a mechanism similar to the one explained in section 4. The
reduction relations are extended preserving analyticity in $\alpha $. Any
new coupling $\overline{\lambda }_{\overline{n}\ell }$ introduced due to
violations of (\ref{invf}) disappears in the limit $\alpha \rightarrow
\alpha ^{*}$ if $Q_{\overline{n}}^{*}\in \mathbb{N}_{+}$ and survives if $%
Q_{n}^{*}=0$. Indeed, as noted above, the invertibility conditions become
just $Q_{n}^{*}\neq 0$ in this limit.

If $Q_{\overline{n}}^{*}$ is a negative integer, then (\ref{solv}) is
meromorphic around the interacting fixed point for $\xi _{\overline{n}}\neq
0 $. Then an arbitrary parameter can be introduced at the price of relaxing
analyticity to perturbative meromorphy. This case is analogous to the one
discussed in section 4, see formula (\ref{more}). The invertibility
conditions, the reduction relations for $n>\overline{n}$, and the effective
Planck mass, are modified following the instructions given in section 4.

\bigskip

Finally, observe that perturbative meromorphy around the free fixed point
and analyticity around the interacting fixed point do not hold at the same
time, in general. This fact can be easily proved integrating (\ref{sola})
exactly in the leading approximation $B(\alpha )=1$,\thinspace $\delta
_{n}(\alpha )=\delta _{n}^{(1)}/\alpha ^{n-2}$, $\gamma _{n}(\alpha )=\gamma
_{n}^{(1)}\alpha $, etc. Indeed, there is no reason why the values of $\xi
_{n}$ that ensure analyticity around the interacting fixed point should
coincide with the values of $\xi _{n}$ that ensure perturbative meromorphy
around the free fixed point. This property is a bit disappointing, but in
the realm of non-renormalizable theories there is no physical reason to
require a nice high-energy limit, so only the IR fixed point matters, free
or interacting.

\section{Applications}

\setcounter{equation}{0}

Some examples are useful to illustrare the arguments of the previous
sections. I consider the Pauli deformation of massless QCD and quantum
gravity coupled with matter in three spacetime dimensions. I also comment on
the difficulties of four-dimensional quantum gravity.

\bigskip

\noindent \textbf{Pauli deformation of massless QCD}

\noindent As an example, consider the conformal window of massless QCD, 
\[
\mathcal{L}=\frac{1}{4\alpha }(F_{\mu \nu }^{a})^{2}+\overline{\psi }%
D\!\!\!\!\slash\psi , 
\]
with $N_{c}$ colors and $N_{f}$ flavors in the fundamental representation,
in the limit where $N_{f},N_{c}$ are large but $N_{f}/N_{c}\lesssim 11/2$.
The UV-fixed point is free, while the IR\ fixed point is interacting, but
weakly coupled, so it can be reached perturbatively. The beta function reads 
\begin{equation}
\beta _{\alpha }=-\frac{\Delta N_{c}}{24\pi ^{2}}\alpha ^{2}+\frac{%
25N_{c}^{2}}{(4\pi )^{4}}\alpha ^{3}+\alpha \sum_{n=3}^{\infty }c_{n}\left(
\alpha N_{c}\right) ^{n},  \label{beta2}
\end{equation}
where $\Delta \equiv 11-2N_{f}/N_{c},$ $0<\Delta \ll 1$ and the $c_{n}$s are
unspecified numerical coefficients. The first two contributions of the beta
function have opposite signs and the first contribution is arbitrarily
small. This ensures that, expanding in powers of $\Delta $, the beta
function has a second zero for 
\begin{equation}
\frac{\alpha _{*}N_{c}}{16\pi ^{2}}=\frac{2}{75}\Delta +\mathcal{O}(\Delta
^{2}),  \label{zero}
\end{equation}
which is the IR\ fixed point of the RG\ flow.

The Pauli deformation \cite{pap3} is the irrelevant deformation with head 
\[
\lambda _{1}~F_{\mu \nu }^{a}~\overline{\psi }T^{a}\sigma _{\mu \nu }\psi 
\]
and has level one. The queue begins with operators of level two. There are
10 four-fermion operators and the $F^{3}$ term 
\[
\frac{\lambda _{2}}{3!}~f^{abc}F_{\mu \nu }^{a}F_{\nu \rho }^{b}F_{\rho \mu
}^{c}. 
\]
For simplicity, I consider only this level-2 term, because the argument is
completely general and the extension to four-fermion operators is
straightforward. The one-loop beta functions are \cite{pap3} 
\[
\beta _{\lambda _{1}}=\frac{\alpha N_{c}}{16\pi ^{2}}\lambda _{1},\qquad
\beta _{\lambda _{2}}=\frac{3N_{c}\alpha }{4\pi ^{2}}\lambda _{2}-\frac{N_{f}%
}{4\pi ^{2}}\lambda _{1}^{2}, 
\]
so in this case 
\[
Q_{2}=\frac{\gamma _{2}^{(1)}-2\gamma _{1}^{(1)}}{\beta _{\alpha }^{(1)}}=-%
\frac{15}{\Delta }=-Q_{2}^{*},\qquad \lambda _{2}=\overline{f}_{2}(\alpha
)\lambda _{1}^{2}=\frac{11}{5\alpha }\lambda _{1}^{2}\left( 1+\mathcal{O}%
(\alpha ,\Delta )\right) . 
\]
Similar formulas hold for $Q_{n}$, $n>2$. Note that in this approximation $%
Q_{n}$ and the $Q_{n}^{*}$ are related in a simple way, namely $%
Q_{n}=-Q_{n}^{*}$. This happens because the interacting fixed point is
weakly coupled.

The invertibility conditions (\ref{invodd}) and (\ref{invf}) for $n=2$ read 
\begin{equation}
1-\frac{15}{\Delta }\notin \mathbb{N},\qquad \frac{15}{\Delta }\notin %
\mathbb{N},  \label{condarun}
\end{equation}
around the free-field limit and the interacting fixed point, respectively.
Similar conditions are expected for $n>2$: 
\begin{equation}
Q_{n}+q_{n}=\frac{b_{n}(\Delta )}{\Delta }\notin \mathbb{N},\qquad Q_{n}^{*}=%
\frac{b_{n}^{\prime }(\Delta )}{\Delta }\notin \mathbb{N},  \label{condarun2}
\end{equation}
$q_{n}$ being $\Delta $-independent quantities depending on the level and
the number of legs of $\mathcal{O}_{n\ell }(\varphi )$, $b_{n}(\Delta )$ and 
$b_{n}^{\prime }(\Delta )$ being rational numbers with smooth $\Delta
\rightarrow 0$ limits $b_{n}(0)=-b_{n}^{\prime }(0)$. A non-trivial
renormalization mixing can only make the invertibility conditions less
restrictive, so I proceed assuming the worst case, which is (\ref{condarun2}%
). Since $\Delta $ is rational and arbitrarily small, while $\Delta $ tends
to zero the numbers $Q_{n}+n-1$ and $Q_{n}^{*}$ cross, among the others,
also natural integer values and so violate the conditions (\ref{condarun2}).
However, if $b_{n}(0)\neq 0$ ($b_{n}^{\prime }(0)\neq 0$), when $%
b_{n}(\Delta )/\Delta \in \mathbb{N}$ ($b_{n}^{\prime }(\Delta )/\Delta \in %
\mathbb{N}$), the new parameters appear at orders $b_{n}(\Delta )/\Delta $ ($%
b_{n}^{\prime }(\Delta )/\Delta $) that are arbitrarily high for $\Delta \ll
1$. Therefore, the effects of the violations of (\ref{condarun2}) are
negligible in the limit discussed here, and the Pauli deformation is
determined unambiguously under the sole conditions $b_{n}(0)\neq 0$, $%
b_{n}^{\prime }(0)\neq 0$, which are equivalent to $\gamma _{n\ell
}^{(1)}\neq n\gamma _{\ell }^{(1)}$ at $\Delta =0$.

Finally, in the limit where the momenta and $1/\lambda _{1}$ are much
smaller than the dynamical scale $\mu $ the deformed theory tends to the
quasi-finite theory studied in ref. \cite{pap3}. Indeed, using (\ref{beha})
it is immediately found that 
\[
\overline{f}_{2}(\alpha _{*})=\frac{165N_{c}}{32\Delta \pi ^{2}}, 
\]
in agreement with \cite{pap3}. Observe that the conditions for the existence
of the irrelevant deformation of the RG flow (namely $\gamma _{n\ell
}^{(1)}\neq n\gamma _{\ell }^{(1)}$ at $\Delta =0$) coincide with the
conditions for the existence of the quasi-finite deformation of the IR fixed
point.

\bigskip

\noindent \textbf{Three-dimensional quantum gravity coupled with matter}

\noindent The results of this paper can be applied also to three-dimensional
quantum gravity coupled with running matter, and generalize the results of
ref. \cite{pap2}, where the matter sector was assumed to be conformal.

If the invertibility conditions are fulfilled, the theory is unique. If new
parameters appear along the way, but they are finitely many, the theory is
still predictive. Finally, if infinitely many parameters are turned on by
the infinite reduction, they are generically expected to appear at
increasingly high orders. Then the predictivity of the theory is of the type
discussed in section 4: it is still possible to make calculations up to very
high orders with a relatively small number of couplings.

\bigskip

\noindent \textbf{Four-dimensional quantum gravity}

\noindent Applications to four-dimensional quantum gravity, instead, demand
further insight. The renormalizable subsector of gravity is not interacting,
so the infinite reduction does not apply. Equivalently, the effective Planck
mass $M_{P\text{eff}}$ is zero, so the perturbative regime $E\ll M_{P\text{%
eff}}$ is empty. The reason why three-dimensional quantum gravity is
exceptional is that, although the renormalizable subsector of gravity is
free, all irrelevant operators constructed with the Riemann tensor and their
derivatives are trivial, in the sense that they can be converted into matter
operators using the field equations. Then, to have a non-trivial effective
Planck mass it is sufficient to have an interacting matter sector.

\section{Irrelevant deformations with several marginal running couplings}

\setcounter{equation}{0}

To complete the investigation of this paper, I study the irrelevant
deformations of running renormalizable theories $\mathcal{R}$ containing
more than one independent marginal coupling. The purpose is to show that the
infinite reduction is free of some difficulties that are present in
Zimmermann's analytic reduction (see Appendix A for details) and better
appreciate some other properties.

I study the behavior of the reduction relations in a neighborhood of the
free fixed point, focusing on the leading-log approximation, for which the
one-loop coefficients of the beta functions and anomalous dimensions
suffice. Consider a renormalizable theory with couplings $\alpha _{1}$ and $%
\alpha _{2}$ and one-loop beta functions 
\[
\beta _{\alpha _{1}}=\beta _{1}\alpha _{1}^{2},\qquad \beta _{\alpha
_{2}}=a\alpha _{1}^{2}+b\alpha _{1}\alpha _{2}+c\alpha _{2}^{2}, 
\]
where $\beta _{1},a,b,c$ are unspecified numerical factors. For intermediate
purposes, it is useful to ``reduce'' the marginal sector to a unique running
constant, say $\alpha _{1}$, plus a finite arbitrary parameter $c_{1}$,
following Zimmermann's method. Solve the RG\ consistency equations 
\[
\frac{\mathrm{d}\alpha _{2}}{\mathrm{d}\alpha _{1}}=\frac{\beta _{\alpha
_{2}}(\alpha _{1},\alpha _{2})}{\beta _{\alpha _{1}}(\alpha _{1},\alpha _{2})%
}. 
\]
The solution reads 
\begin{equation}
\widetilde{\alpha }_{2}(\alpha _{1},c_{1})=-\frac{\alpha _{1}}{2c}\left[
b-\beta _{1}+s\frac{1+(\alpha _{1}/c_{1})^{-s/\beta _{1}}}{1-(\alpha
_{1}/c_{1})^{-s/\beta _{1}}}\right] ,  \label{alfa2tilde}
\end{equation}
where $s=\sqrt{(b-\beta _{1})^{2}-4ac}$. The quantity $s$ can be complex,
but this does not cause problems here, because (\ref{alfa2tilde}) is used
only for intermediate purposes. In the end $c_{1}$ is eliminated in favor of 
$\alpha _{2}$ using the inverse of (\ref{alfa2tilde}): 
\begin{equation}
c_{1}=\alpha _{1}z^{-\beta _{1}/s},\qquad \text{where }z=\frac{2c\alpha
_{2}+\alpha _{1}(b-\beta _{1}-s)}{2c\alpha _{2}+\alpha _{1}(b-\beta _{1}+s)}.
\label{c1}
\end{equation}
Observe that the function $\alpha _{1}z^{-\beta _{1}/s}$ is constant along
the RG flow. The noticeable modular combination $z$ plays an important role
throughout the discussion.

Consider an irrelevant deformation of level $\ell $, with coupling $\lambda
_{\ell }$. The first term of the queue is multiplied by the coupling $%
\lambda _{2\ell }$. For small $\alpha _{1,2}$, the lowest-order beta
functions of $\lambda _{\ell }$ and $\lambda _{2\ell }$ have generically the
forms 
\begin{equation}
\beta _{\lambda _{\ell }}=\lambda _{\ell }(d\alpha _{1}+e\alpha _{2}),\qquad
\beta _{\lambda _{2\ell }}=\lambda _{2\ell }(f\alpha _{1}+g\alpha
_{2})+h\lambda _{\ell }^{2},  \label{duebeta}
\end{equation}
where $d,e,f,g,h$ are unspecified numerical factors. Search for a reduction
relation of the form 
\[
\lambda _{2\ell }=f_{2}(\alpha _{1},\alpha _{2})\lambda _{\ell }^{2}. 
\]
Differentiating this expression and using (\ref{duebeta}), the equation
obeyed by $f_{2}$ reads 
\begin{equation}
\frac{\mathrm{d}f_{2}}{\mathrm{d}\ln \mu }=h-2f_{2}~(\widetilde{d}\alpha
_{1}+\widetilde{e}\alpha _{2}),  \label{RGF}
\end{equation}
where $\widetilde{d}=d-f/2$ and $\widetilde{e}=e-g/2$. Now, (\ref{RGF}) is
one differential equation for a function of two variables, so the most
general solution contains an arbitrary function of one variable (see below).
Call 
\begin{equation}
\widetilde{f}_{2}(\alpha _{1},c_{1})=f_{2}\left( \alpha _{1},\widetilde{%
\alpha }_{2}(\alpha _{1},c_{1})\right)  \label{Ftilde}
\end{equation}
the solution of the equation 
\begin{equation}
\beta _{1}\alpha _{1}^{2}\frac{\mathrm{d}\widetilde{f}_{2}(\alpha _{1},c_{1})%
}{\mathrm{d}\alpha _{1}}+2\widetilde{f}_{2}(\alpha _{1},c_{1})\left( 
\widetilde{d}\alpha _{1}+\widetilde{e}\widetilde{\alpha }_{2}(\alpha
_{1},c_{1})\right) =h,  \label{RGFtilde}
\end{equation}
which is obtained inserting (\ref{alfa2tilde}) into (\ref{RGF}). The
solution of (\ref{Ftilde}) depends on $c_{1}$ and a further arbitrary
constant $c_{2}$. Eliminating $c_{1}$ with the help of (\ref{c1}), the
solution reads 
\begin{equation}
f_{2}(\alpha _{1},\alpha _{2})=\overline{f}_{2}(\alpha _{1},\alpha
_{2})+c_{2}\overline{s}_{2}(\alpha _{1},\alpha _{2}),  \label{f1}
\end{equation}
where 
\begin{equation}
\overline{f}_{2}(\alpha _{1},\alpha _{2})=\frac{2h~_{2}F_{1}[1,\gamma -2%
\widetilde{e}/c,\gamma ,z]}{\left( 2c\alpha _{2}+\alpha _{1}(s+b-\beta
_{1})\right) (\gamma -1)},\qquad \gamma =\frac{1}{cs}\left( c(2\widetilde{d}%
-\beta _{1}+s)+\widetilde{e}(s-b+\beta _{1})\right) ,  \label{f2}
\end{equation}
and 
\begin{equation}
\overline{s}_{2}(\alpha _{1},\alpha _{2})=z^{-\delta }~(2c\alpha _{2}+\alpha
_{1}(s+b-\beta _{1}))^{-2\widetilde{e}/c},\qquad \delta =\frac{\widetilde{e}%
(s-b-\beta _{1})+2c\widetilde{d}}{cs}.  \label{f3}
\end{equation}
Now, $c_{2}$ is an arbitrary constant of the RG equation (\ref{RGF}). This
means that $c_{2}$ is constant only along the RG flow, but can otherwise
depend on $\alpha _{1}$ and $\alpha _{2}$. The function of $\alpha _{1}$ and 
$\alpha _{2}$ that is constant along the RG flow is given in eq. (\ref{c1}),
so $c_{2}$ is an arbitrary function $k_{2}$ of $\alpha _{1}z^{-\beta _{1}/s}$%
. In conclusion, the most general solution of (\ref{RGF}) reads 
\[
f_{2}(\alpha _{1},\alpha _{2})=\overline{f}_{2}(\alpha _{1},\alpha
_{2})+k_{2}(\alpha _{1}z^{-\beta _{1}/s})~\overline{s}_{2}(\alpha
_{1},\alpha _{2}). 
\]

The remarkable points are $z=0,1,\infty $, i.e. 
\begin{equation}
\alpha _{2}+\frac{\alpha _{1}}{2c}(b-\beta _{1}-s)=0,\qquad \alpha
_{1}=0,\qquad \alpha _{2}+\frac{\alpha _{1}}{2c}(b-\beta _{1}+s)=0,
\label{lines}
\end{equation}
respectively. These are the combinations of couplings that vanish together
with their own beta functions at the leading-log level. Along these lines a
subsector of the theory is practically at a fixed point, in the given
approximation. Therefore, the reduction should be perturbatively
meromorphic, or analytic, in the neighborhood of such lines. However,
formulas (\ref{f1}), (\ref{f2}) and (\ref{f3}) show that perturbative
meromorphy can be imposed only in the neighborhood of one line (\ref{lines})
at a time, not around all of them contemporarily. The situation is similar
to the one discussed in section 6, where it was observed that perturbative
meromorphy around the free-field limit and analyticity around the
interacting fixed point mutually exclude each other. Once the line of
perturbative meromorphy is chosen, the function $k_{2}$ is uniquely
determined: in the order (\ref{lines}), the results are 
\begin{eqnarray*}
\overline{f}_{2}(\alpha _{1},\alpha _{2}) &=&\frac{2h~_{2}F_{1}[1,\gamma -2%
\widetilde{e}/c,\gamma ,z]}{\left( 2c\alpha _{2}+\alpha _{1}(b-\beta
_{1}+s)\right) (\gamma -1)},\qquad 1-\gamma \notin \mathbb{N}, \\
\overline{f}_{2}^{\prime }(\alpha _{1},\alpha _{2}) &=&\frac{%
2h~_{2}F_{1}[1,\gamma -2\widetilde{e}/c,2-2\widetilde{e}/c,1-z]}{\left(
2c\alpha _{2}+\alpha _{1}(b-\beta _{1}+s)\right) (2\widetilde{e}/c-1)}%
,\qquad 2\frac{\widetilde{e}}{c}-1\notin \mathbb{N}, \\
\overline{f}_{2}^{\prime \prime }(\alpha _{1},\alpha _{2}) &=&-\frac{%
2h~_{2}F_{1}[1,2-\gamma ,2-\gamma +2\widetilde{e}/c,1/z]}{\left( 2c\alpha
_{2}+\alpha _{1}(b-\beta _{1}-s)\right) (\gamma -2\widetilde{e}/c-1)},\qquad
\gamma -2\frac{\widetilde{e}}{c}-1\notin \mathbb{N}.
\end{eqnarray*}
To the right the respective existence conditions are reported.

The other terms of the queue are worked out similarly. As before, the
invertibility conditions involve only scheme-independent coefficients and
new independent couplings can sporadically appear at high orders.

\section{Conclusions}

\setcounter{equation}{0}

I have studied methods to classify the non-renormalizable interactions and
criteria to remove the infinite arbitrariness of non-renormalizable
theories, taking inspiration from recent constructions of finite and
quasi-finite irrelevant deformations of interacting conformal field
theories. I have considered non-renormalizable theories whose renormalizable
subsector $\mathcal{R}$ is fully interacting, running, with one or more
marginal couplings. Relevant couplings can be added perturbatively to the
constructions of this paper.

An irrelevant deformation is made of a head and a queue that runs coherently
with the head. The head is the lowest-level irrelevant term, multiplied by
an independent coupling $\lambda _{\ell }$. The queue is made of an infinite
number of irrelevant terms with higher dimensionalities in units of mass.
``Reduction'' relations express the couplings of the queue as functions of $%
\lambda _{\ell }$ and the marginal couplings $\alpha $ of $\mathcal{R}$. The
reduction relations are polynomial in $\lambda _{\ell }$. The $\alpha $
-dependence is determined by consistency with the renormalization group and
one of the following scheme-independent prescriptions: $i$) perturbative
meromorphy around a free fixed point of $\mathcal{R}$, or $ii$) analyticity
around an interacting fixed point of $\mathcal{R}$. In general, it is not
possible to have both at the same time. The infinite reduction works when
certain invertibility conditions are fulfilled. In the case of violations,
new independent couplings $\overline{\lambda }_{\text{new}}$ are introduced
along the way. The divergences of a theory reduced with these criteria are
reabsorbed into renormalization constants for $\alpha $, $\lambda _{\ell }$
and eventually $\overline{\lambda }_{\text{new}}$, plus field redefinitions.

With prescription $i$) the number of independent couplings remains finite or
grows together with the order of the expansion. It remains finite if the
irrelevant operators have a sufficiently non-trivial renormalization mixing.
When the number of couplings grows together with the order of the expansion,
the growth is in general so slow that a reasonably small number of couplings
are sufficient to make predictions up to very high orders. With prescription 
$ii$) the number of couplings generically remains finite.

The infinite reduction does not determine which non-renormalizable
interactions are switched on and off in nature, but is the basic tool to
classify the non-renormalizable interactions and address the search for
selective criteria. In my opinion the theories costructed with the
infinite-reduction prescription are as fundamental as the usual
renormalizable theories.

\vskip 25truept \noindent {\Large \textbf{Acknowledgments}}

\vskip 15truept \noindent

I thank the CBPF\ (Centro Brasileiro de Pesquisas Fisicas) of Rio de Janeiro
for hospitality during the early stage of this work and M. Halat for useful
discussions.

\vskip 25truept \noindent {\Large \textbf{A\ \ Appendix: Zimmermann's
reduction of couplings}}\vskip 15truept

\renewcommand{\theequation}{A.\arabic{equation}} \setcounter{equation}{0}

\noindent In this section I\ review the main properties of Zimmermann's
``reduction of couplings'' \cite{zimme}. I also describe some difficulties
of the analytic prescription and emphasize the properties of perturbative
meromorphy. It is convenient to have a concrete example in mind, such as
massless scalar electrodynamics, 
\begin{equation}
\mathcal{L}=\frac{1}{4\alpha }F_{\mu \nu }^{2}+|D_{\mu }\varphi |^{2}+\frac{%
\lambda }{4}(\overline{\varphi }\varphi )^{2},  \label{nonsusy}
\end{equation}
where $D_{\mu }\varphi =\partial _{\mu }\varphi +iA_{\mu }\varphi $. The
reduction is a function $\lambda (\alpha )$ that relates the two couplings.
Consistency with the renormalization group gives the differential equation 
\begin{equation}
\frac{\mathrm{d}\lambda (\alpha )}{\mathrm{d}\alpha }=\frac{\beta _{\lambda
}(\lambda (\alpha ),\alpha )}{\beta _{\alpha }(\lambda (\alpha ),\alpha )},
\label{zimmesp}
\end{equation}
that determines the solution $\lambda (\alpha )$ up to an arbitrary constant 
$\overline{\lambda }$, the initial condition. The structures of the beta
functions are 
\begin{eqnarray}
\beta _{\alpha } &=&\alpha \left( \beta _{1}\alpha +\beta _{21}\lambda
^{2}+\beta _{22}\lambda \alpha +\beta _{23}\alpha ^{2}+\cdots \right) , 
\nonumber \\
\beta _{\lambda } &=&a_{1}\alpha ^{2}+a_{2}\lambda \alpha +a_{3}\lambda
^{2}+b_{1}\lambda ^{3}+b_{2}\lambda ^{2}\alpha +b_{3}\lambda \alpha
^{2}+b_{4}\alpha ^{3}+\cdots .  \label{betaso}
\end{eqnarray}
Assume $\beta _{1}\neq 0$. If 
\begin{equation}
\Delta \equiv (a_{2}-\beta _{1})^{2}-4a_{1}a_{3}\geq 0,  \label{co1}
\end{equation}
and 
\begin{equation}
r_{\pm }\equiv \pm \frac{s}{\beta _{1}}-1\notin \mathbb{N},  \label{co2}
\end{equation}
where $s$ is the positive square root of $\Delta $, then the equations (\ref
{zimmesp}) are solved by the expansions 
\begin{equation}
\lambda _{\pm }(\alpha ,d)=\sum_{k=1}^{\infty }c_{\pm k}\alpha
^{k}+\sum_{m,n=1}^{\infty }d_{\pm mn}d^{n}\alpha ^{m\pm ns/\beta _{1}},
\label{expa}
\end{equation}
where 
\[
c_{\pm 1}=\frac{1}{2a_{3}}\left( \beta _{1}-a_{2}\pm s\right) ,\qquad d_{\pm
11}=1, 
\]
and the coefficients $c_{\pm k}$, $d_{\pm mn}$ are unambiguous calculable
numbers, while $d$ is an arbitrary parameter. If $\beta _{1}>0$ the
meaningful expansions are $\lambda _{+}(\alpha ,d)$ and $\lambda _{-}(\alpha
,0)$, if $\beta _{1}<0$ they are $\lambda _{+}(\alpha ,0)$ and $\lambda
_{-}(\alpha ,d)$.

The condition (\ref{co1}) follows from the reality of $\lambda $ in the
expansion (\ref{expa}), while (\ref{co2}) are derived inserting (\ref{expa})
in (\ref{betaso}) and (\ref{zimmesp}). The coefficients $c_{\pm k}$ have
expressions 
\begin{equation}
c_{\pm k}=\frac{P_{\pm k}(c_{\pm 1},a,b,\beta ,\ldots )}{\prod_{j=2}^{k}%
\left[ (1-j)\beta _{1}\pm s\right] },\qquad k>1,  \label{col}
\end{equation}
where $P_{\pm k}$ are polynomials that in general do not vanish when the
denominator vanishes.

Clearly, the expansions $\lambda _{\pm }(\alpha ,d)$ (\ref{expa}) are just
different expansions of the same function $\lambda (\alpha ,d)$, because for
every value of $d$ the solution is unique.

The condition (\ref{co1}) is quite restrictive, and excludes a good amount
of models. On the other hand, when (\ref{co1}) holds, $s$ is generically an
irrational number and (\ref{co2}) is automatically satisfied. Therefore in
Zimmermann's reduction the crucial existence condition is (\ref{co1}).

As long as $d$ is arbitrary, there is no true reduction. The $d$-ambiguity
can be eliminated demanding that the reduction be analytic. When (\ref{co1})
and (\ref{co2}) hold, both expansions $\lambda _{\pm }(\alpha ,d)$ are
generically non-analytic at $d\neq 0$. Therefore analyticity implies $d=0$
and gives two distinct unambiguous solutions 
\begin{equation}
\lambda _{\pm }(\alpha )\equiv \lambda _{\pm }(\alpha ,0)=\sum_{n=1}^{\infty
}c_{_{\pm }k}\alpha ^{k}.  \label{tuttiluppi}
\end{equation}

On the other hand, if (\ref{co1}) holds, but (\ref{co2}) does not, then
analyticity is violated by logarithms, which signal the presence of the
other independent coupling. In this case the reduction is ineffective.

\bigskip

Now I compare Zimmermann's reduction with the infinite reduction. If the
renormalizable aubsector $\mathcal{R}$ does not contain relevant couplings,
the beta function of an irrelevant coupling $\lambda $ is linear in $\lambda 
$. Then the infinite reduction has existence conditions of type (\ref{co2}),
one for every term of the queue, but no existence condition of type (\ref
{co1}). This is a lucky situation, since it would be hopeless to satisfy
infinitely many reality conditions such as (\ref{co1}). On the other hand,
the conditions of type (\ref{co2}) become (\ref{conditions}) and involve
only rational numbers. It is not unfrequent that some of these rational
numbers coincide with natural numbers $\overline{k}$. For every such
``coincidence'' a new independent coupling is introduced at order $\overline{%
k}$.

\bigskip

\noindent \textbf{Difficulties of Zimmermann's approach with two or more
reduced couplings}

\noindent Consider a generic renormalizable theory with three marginal
couplings, $\alpha _{1},\alpha _{2}$ and $\lambda $, with one-loop beta
functions 
\begin{eqnarray*}
\beta _{\alpha _{1}} &=&\beta _{1}\alpha _{1}^{2},\qquad \beta _{\alpha
_{2}}=a\alpha _{1}^{2}+b\alpha _{1}\alpha _{2}+c\alpha _{2}^{2}, \\
\beta _{\lambda } &=&f\alpha _{1}^{2}+g\alpha _{1}\alpha _{2}+h\alpha
_{2}^{2}+\lambda (d\alpha _{1}+e\alpha _{2})+l\lambda ^{2},
\end{eqnarray*}
and seek for an analytic reduction 
\begin{equation}
\lambda (\alpha _{1},\alpha _{2})=c_{1}\alpha _{1}+d_{1}\alpha
_{2}+c_{2}\alpha _{1}^{2}+d_{2}\alpha _{1}\alpha _{2}+e_{2}\alpha
_{2}^{2}+\cdots ,  \label{nsion}
\end{equation}
leaving two independent couplings and eliminating the third one.
Differentiating (\ref{nsion}) and matching the coefficients of $\alpha
_{1}^{2}$, $\alpha _{1}\alpha _{2}$ and $\alpha _{2}^{2}$ with $\beta
_{\lambda }$, the following equations are obtained: 
\begin{eqnarray*}
c_{1}\beta _{1}+d_{1}a &=&f+c_{1}d+lc_{1}^{2}, \\
d_{1}b &=&g+c_{1}e+d_{1}d+2lc_{1}d_{1}, \\
d_{1}c &=&h+d_{1}e+ld_{1}^{2}.
\end{eqnarray*}
These are three (generically independent) equations for the two unknowns $%
c_{1}$ and $d_{1}$. The mismatch between the number of unknows and the
number of equations has the following explanation. The expansion (\ref{nsion}%
) is made of a sum of polynomials of degrees $n=1,2,\ldots $ in $\alpha _{1}$
and $\alpha _{2}$. The polynomial of degree $n$ contributing to $\lambda
(\alpha _{1},\alpha _{2})$ contains $n+1$ monomials and therefore $n+1$
unknown coefficients. After differentiation, these unknowns contribute to
polynomials of higher orders, at least $n+2$, in the RG\ consistency
conditions, thereby they appear in at least $n+2$ equations. Therefore the
problem has, in general, no solutions \cite{wilson}.

This means that the analyticity requirement is too strong. The problems are
avoided as explained in section 8.

\vskip 25truept \noindent {\Large \textbf{B\ \ Appendix: Perturbative
meromorphy and infinite reduction}}

\vskip 15truept

\renewcommand{\theequation}{B.\arabic{equation}} \setcounter{equation}{0}

\noindent In this appendix I define the notion of perturbative meromorphy
and study some of its properties. A function $f(\lambda ,\alpha )$ is said
to be perturbatively meromorphic in $\alpha $ with respect to $\lambda $ if
it is analytic in $\lambda $ and admits an expansion 
\[
f(\lambda ,\alpha )=g(\alpha )+\sum_{n=1}^{\infty }c_{n}(\alpha )\lambda
^{n}, 
\]
such that the functions $g(\alpha ),c_{n}(\alpha )$ are meromorphic in $%
\alpha $ and the $c_{n}(\alpha )$s have at most poles of order $pn-q$, where 
$p$ and $q$ are non-negative. Assume for definiteness that the poles are in $%
\alpha =0$. Then it is clear that if $\lambda _{\text{eff}}=\lambda \alpha
^{-p}$ the function 
\[
\overline{f}(\lambda _{\text{eff}},\alpha )=f(\lambda _{\text{eff}}\alpha
^{p},\alpha )=g(\alpha )+\alpha ^{q}\sum_{n=0}^{\infty }\overline{c}%
_{n}(\alpha )\lambda _{\text{eff}}^{n}, 
\]
is analytic in $\lambda _{\text{eff}}$ and meromorphic in $\alpha $, since
the $\overline{c}_{n}(\alpha )$s are regular.

For example, renormalizable quantum field theory is perturbatively
meromorphic in $\varepsilon $ with respect to $\hbar $ or the marginal
couplings $\alpha $, in the sense that the renormalization constants admit
expansions of the form 
\[
\sum_{L=0}^{\infty }\sum_{k=0}^{L}c_{L,k}\left( \frac{1}{\varepsilon }%
\right) ^{k}\alpha ^{L}+\text{evanescent}=\sum_{L=0}^{\infty
}c_{L}(\varepsilon )\alpha ^{L} 
\]
where $L$ is the number of loops and $c_{L}(\varepsilon )$ has at most a
pole of order $L$. Here $p=1$ and $q=0$.

The infinite reduction is perturbatively meromorphic in the marginal
couplings $\alpha $ with respect to the irrelevant couplings $\lambda _{\ell
}$. When $\alpha \sim 0$%
\begin{eqnarray}
\mathcal{L}[\varphi ] &\sim &\mathcal{L}_{\mathcal{R}}[\varphi ,\alpha
]+\lambda _{\ell }\mathcal{O}_{\ell }(\varphi )+\sum_{n=2}^{\infty }\frac{%
c_{n}\alpha ^{k_{n}}}{\alpha ^{pn-q}}\lambda _{\ell }^{n}~\mathcal{O}_{n\ell
}(\varphi )  \label{ref1} \\
&=&\mathcal{L}_{\mathcal{R}}[\varphi ,\alpha ]+\alpha ^{p}\lambda _{\ell 
\mathrm{eff}}\mathcal{O}_{\ell }(\varphi )+\alpha ^{q}\sum_{n=2}^{\infty
}c_{n}\alpha ^{k_{n}}\lambda _{\ell \mathrm{eff}}^{n}~\mathcal{O}_{n\ell
}(\varphi ),  \label{ref2}
\end{eqnarray}
where $\lambda _{\ell \mathrm{eff}}=\lambda _{\ell }\alpha ^{-p}$, $c_{n}$
are constants and $k_{n}$ are non-negative integers. Normally, in the
absence of three-leg vertices $p=q\geq 1$. In the presence of three-leg
vertices the elementary marginal coupling is $g=\alpha ^{1/2}$, which is
equivalent to say $p\geq q=1/2$. The number $q$ is the \textit{order} of the
irrelevant deformation. The order is always positive, which emphasizes that
it is necessary to have an interacting renormalizable subsector to build the
irrelevant deformations.

Observe that marginal and irrelevant deformations do not commute. It is
possible to switch the irrelevant deformation off, keeping $\alpha \neq 0$,
but it is impossible to switch the marginal deformation off keeping a
non-trivial irrelevant sector in the limit $\alpha \rightarrow 0$. This is
also the meaning of perturbative meromorphy: in some sense, the true head of
the irrelevant deformation is the operator of infinite dimensionality.

Now I prove some theorems that are useful in the paper. Observe that for
each level $n$ that fulfills the invertibility conditions, (\ref{conditions}%
) or their appropriate generalizations, the differential operator 
\begin{equation}
D_{n}\equiv \beta _{\alpha }\frac{\mathrm{d}}{\mathrm{d}\alpha }-\gamma
_{n\ell }(\alpha )+n\gamma _{\ell }(\alpha )  \label{dope}
\end{equation}
appearing in the RG\ consistency conditions (\ref{rg}) is $\mathcal{O}%
(\alpha )$ and its order $\mathcal{O}(\alpha )$ is non-vanishing. Then (\ref
{dope}) can be freely inverted and $D_{n}^{-1}=\mathcal{O}(\alpha ^{-1})$.

\bigskip

\textbf{Theorem B1.} Suppose that $q\geq 1$, $2p\geq q+1$, that the
renormalization structures (\ref{ref1}),(\ref{ref2}) are stable under
renormalization up to the level $\overline{n}$ and that each level $>%
\overline{n}$ fulfills the invertibility conditions. Then the structures (%
\ref{ref1}),(\ref{ref2}) are stable under renormalization.

\noindent \textbf{Proof.} Consider (\ref{ref2}) and the beta functions (\ref
{reft}) $\beta _{n\ell }=\gamma _{n\ell }\lambda _{n\ell }+\delta _{n\ell }$%
, $n>\overline{n}$. Since $\delta _{n\ell }$ is at least quadratic in the
irrelevant couplings and in the notation (\ref{ref2}) each of coupling
carries at least a power $\alpha ^{s}$, $s=\min (p,q)$, then $\delta _{n\ell
}\sim \alpha ^{2s}\lambda _{\ell \text{eff}}^{n}$ at worst, so, using $%
D_{n}^{-1}=\mathcal{O}(\alpha ^{-1})$, $\lambda _{n\ell }\sim \alpha
^{2s-1}\lambda _{\ell \text{eff}}^{n}=\alpha ^{q}\alpha ^{2s-q-1}\lambda
_{\ell \text{eff}}^{n}$, which is compatible with (\ref{ref2}), since $%
2s-q-1\geq 0$.

\bigskip

\textbf{Theorem B2.} Suppose that $\lambda _{k\ell }$ behaves at worst as 
\begin{equation}
\lambda _{k\ell }\sim \frac{\lambda _{\ell }^{k}}{\alpha ^{q(k-1)}},
\label{itis}
\end{equation}
for $1<k<n$, with $q\geq 1$, and that the invertibility conditions are
fulfilled for $k\geq n$. Then the irrelevant deformation is perturbatively
meromorphic of order $q$.

\noindent \textbf{Proof.} By induction, it is sufficient to prove (\ref{itis}%
) for $k=n$. Consider again the beta functions (\ref{reft}) $\beta _{n\ell
}=\gamma _{n\ell }\lambda _{n\ell }+\delta _{n\ell }$. Since $\delta _{n\ell
}$ depends on the $\lambda _{k\ell }$ with $k<n$, $\delta _{n\ell }\sim
\prod_{k<n}\lambda _{k\ell }{}^{n_{k}}(1+\mathcal{O}(\alpha ))$, with $%
\sum_{k<n}kn_{k}=n$, where $n_{k}$ are non-negative integers. Moreover, $%
m\equiv \sum_{k<n}n_{k}\geq 2$, since $\delta _{n\ell }$ is at least
quadratic. Therefore for small $\alpha $, using $D_{n}^{-1}=\mathcal{O}%
(\alpha ^{-1})$, 
\begin{equation}
\lambda _{n\ell }\sim \frac{\delta _{n\ell }}{\alpha }\sim \frac{\lambda
_{\ell }^{n}}{\alpha }\prod_{k<n}\left( \frac{1}{\alpha ^{q(k-1)}}\right)
^{n_{k}}=\frac{\lambda _{\ell }^{n}}{\alpha ^{q(n-m)+1}}\leq \frac{\lambda
_{\ell }^{n}}{\alpha ^{q(n-1)}}.  \label{prova}
\end{equation}

\bigskip

\textbf{Theorem B3. }Suppose that $\lambda _{k\ell }$ behaves at worst as 
\begin{equation}
\lambda _{k\ell }\sim \frac{\lambda _{\ell }^{k}}{\alpha ^{q(k-1)}},
\label{iu}
\end{equation}
with $q\geq 1$, for $k<\overline{n}$, that $\lambda _{\overline{n}\ell }$
has a more singular behavior 
\begin{equation}
\lambda _{\overline{n}\ell }\sim \frac{\lambda _{\ell }^{\overline{n}}}{%
\alpha ^{q(\overline{n}-1)+\overline{r}}},  \label{iu2}
\end{equation}
with $\overline{r}>0$, and that the invertibility conditions are fulfilled
for $k>\overline{n}$. Then the behavior of $\lambda _{k\ell }$ for generic $%
k $ is at worst 
\begin{equation}
\lambda _{k\ell }\sim \frac{\lambda _{\ell }^{k}}{\alpha ^{q(k-1)+\overline{r%
}[k/\overline{n}]}}  \label{ipo}
\end{equation}
and the multiple irrelevant deformation is perturbatively meromorphic of
order 
\begin{equation}
\overline{q}=q+\left[ \frac{\overline{r}}{\overline{n}-1}\right] _{+},
\label{cu}
\end{equation}
$[x]_{+}$ denoting the minimum integer $\geq x$. Moreover, the associated
renormalization structure (\ref{ref2}) with $p,q\rightarrow $ $\overline{q}$
is stable under renormalization.

\noindent \textbf{Proof.} Formula (\ref{ipo}) certainly holds for $k\leq 
\overline{n}$. By induction, assuming (\ref{ipo}) for $k<n$, with $n>%
\overline{n}$, it is sufficient to prove (\ref{ipo}) for $k=n$. Indeed,
using the same notation as in the proof of Theorem B2, 
\[
\lambda _{n\ell }\sim \frac{\delta _{n\ell }}{\alpha }\sim \frac{\lambda
_{\ell }^{n}}{\alpha }\prod_{k<n}\left( \frac{1}{\alpha ^{q(k-1)+\overline{r%
}[k/\overline{n}]}}\right) ^{n_{k}}=\frac{\lambda _{\ell }^{n}}{\alpha
^{q(n-m)+1+\overline{r}\sum_{k<n}n_{k}[k/\overline{n}]}}\leq \frac{\lambda
_{\ell }^{n}}{\alpha ^{q(n-1)+\overline{r}[n/\overline{n}]}}. 
\]
The last inequality follows from 
\[
q(m-1)\geq 1,\qquad \left[ \frac{n}{\overline{n}}\right] \geq
\sum_{k<n}n_{k}\left[ \frac{k}{\overline{n}}\right] . 
\]

Moreover, (\ref{ipo}) implies 
\begin{equation}
\lambda _{n\ell }\leq \frac{\lambda _{\ell }^{n}}{\alpha ^{\overline{q}(n-1)}%
},  \label{ur}
\end{equation}
for every $n$. Indeed, for $n<\overline{n}$ (\ref{ur}) follows from $%
\overline{q}\geq q$. For $n=\overline{n}$ (\ref{ur}) follows from $%
[x]_{+}\geq x$. For $n>\overline{n}$ (\ref{ur}) follows from 
\[
\overline{r}\left[ \frac{n}{\overline{n}}\right] \leq \overline{r}\frac{n}{%
\overline{n}}<\overline{r}\frac{n-1}{\overline{n}-1}\leq \left[ \frac{%
\overline{r}}{\overline{n}-1}\right] _{+}(n-1). 
\]
So, the irrelevant deformation is perturbatively meromorphic of order $%
\overline{q}$. Finally, since $\overline{q}\geq 1$, by Theorem B1 the
associated renormalization structure is stable.

\end{document}